\documentclass[%
 reprint,
superscriptaddress,
showpacs,preprintnumbers,
 amsmath,amssymb,
 aps,
prb,
letterpaper
]{revtex4-1}

\usepackage{graphicx}
\usepackage{dcolumn}
\usepackage{bm}

\usepackage{amsmath,amsfonts,amssymb,amsthm}
\usepackage{txfonts}
\usepackage{enumerate}
\usepackage{color}
\usepackage{braket}
\usepackage{cases}
\usepackage{comment}

\makeatletter
\renewcommand{\theequation}{\arabic{section}.\arabic{equation}}
\@addtoreset{equation}{section}
\makeatother

\begin{document}

\preprint{APS/123-QED}

\title{Nambu-Goldstone modes propagating along 
topological defects:\\ 
Kelvin and ripple modes from small to large systems}
\author{Daisuke A. Takahashi}\email{daisuke.takahashi.ss@riken.jp}
\affiliation{RIKEN Center for Emergent Matter Science (CEMS), Wako, Saitama 351-0198, Japan}
\affiliation{Research and Education Center for Natural Sciences, Keio University, Hiyoshi 4-1-1, Yokohama, Kanagawa 223-8521, Japan}
\author{Michikazu Kobayashi}
\affiliation{Department of Physics, Kyoto University, Oiwake-cho, Kitashirakawa, Sakyo-ku, Kyoto 606-8502, Japan}
\author{Muneto Nitta}
\affiliation{Research and Education Center for Natural Sciences, Keio University, Hiyoshi 4-1-1, Yokohama, Kanagawa 223-8521, Japan}
\affiliation{Department of Physics, Keio University, Hiyoshi 4-1-1, Yokohama, Kanagawa 223-8521, Japan}

\date{\today}

\begin{abstract}
Nambu-Goldstone modes associated with (topological) defects such as vortices and domain walls in (super)fluids are known to possess quadratic/non-integer dispersion relations in finite/infinite-size systems. 
Here, we report interpolating formulas connecting the dispersion relations in finite- and infinite-size systems for Kelvin modes along a quantum vortex and ripplons on a domain wall in superfluids. 
Our method can provide not only the dispersion relations but also the explicit forms of quasiparticle wavefunctions $(u,v)$. 
We find a complete agreement between 
the analytical formulas and numerical simulations. All these formulas are derived in a fully analytical way, and hence not empirical ones.
We also discuss  common structures in the derivation of these formulas and speculate on the general procedure.
\end{abstract}

\maketitle

\section{Introduction}
In the latter part of the 19th century, Lord Kelvin left many influential works in classical fluid mechanics, and not a few of them form the foundation of this research field today. Among his works are those on the propagation of linear waves in the vicinity of local inhomogeneous structure of fluids, such as ripple modes (capillary waves) along an interface between two fluids \cite{Kelvin1871}, which arise as a by-product of the study of Kelvin-Helmholtz instability (e.g., Ref.~\onlinecite{Chandrasekhar}), and helical motions of vortices, which are now called Kelvin modes \cite{Kelvin1880}. These modes are notable for the point that they have non-integer dispersion relations: while the ripple modes have a fractional dispersion relation $ \epsilon \propto k^{3/2} $ (Ref. \onlinecite{LandauLifshitzfluid}), the Kelvin modes have a logarithmic one $ \epsilon \propto -k^2\log k $. 

In modern physics, these linear waves are also known to emerge in various examples of quantum fluids. 
The Kelvin modes, 
or Kelvons if observed as quantized quasiparticles, 
exist in quantized vortices 
in superfluids \cite{Pitaevskii1961,PhysRev.162.143,RevModPhys.59.87,Donnelly}, Bose-Einstein condensates (BECs) 
of ultracold atomic gases \cite{PhysRevLett.90.100403,PhysRevA.69.043617,PhysRevLett.101.020402}, 
or neutron superfluids in neutron stars, 
having the same dispersion relation with classical fluids 
in the infinite-volume limit. 
Kelvin modes are considered to play 
an important role known as the Kelvin-mode cascade in turbulences, 
including quantum turbulence \cite{Svistunov-Kelvin,Vinen-Kelvin}. 
Thus, understanding Kelvin modes better is  
an important step toward complete characterization of turbulences, 
which remains an unsolved problem 
since the first observation by da Vinci. 
The ripple modes, or ripplons if identified as quasiparticles, emerge on a domain wall\cite{PhysRevLett.81.5718,PhysRevA.58.4836,PhysRevA.78.023624} (DW) 
of a mixture of two kinds of BECs 
and also possess the same dispersion relation with classical fluids in infinite-size systems \cite{PhysRevA.65.033618,PhysRevA.88.043612},  
and the analogous phenomena of the Kelvin-Helmholtz and Rayleigh-Taylor instabilities were also found \cite{PhysRevA.82.063604,PhysRevA.83.043623,PhysRevA.80.063611}. There are also related issues\cite{PhysRevA.89.013617,PhysRevA.90.023612,ArkoRoy}.

Recently, a new insight has been brought to
these gapless modes,  
stimulated by a renewed understanding on 
Nambu-Goldstone modes (NGMs) in non-relativistic systems 
\cite{Nielsen:1975hm,Nambu:2004yia,Watanabe:2011ec,
Watanabe:2012hr,Hidaka:2012ym,Watanabe:2014zza}. 
Both Kelvin modes \cite{Kobayashi01022014,DTMN} 
and ripple modes \cite{PhysRevA.88.043612,DTMN} 
have quadratic dispersion relations, 
$ \epsilon \sim (\log R)k^2 $ and 
$ \epsilon \sim \sqrt{L}k^2 $, 
in finite-size systems with $R$ and $L$ denoting system lengths perpendicular to 
a vortex and DW, respectively. 
These facts are 
consistent with the general argument that 
an  NGM with quadratic dispersion corresponds to two broken symmetries 
\cite{Nambu:2004yia,Watanabe:2011ec,Watanabe:2012hr,Hidaka:2012ym}.
In the limit $R,L\to \infty$, however,
we encounter a difficulty of the divergent coefficient and the correct dispersion laws change to the non-integer ones mentioned above.
How these qualitatively different integer and non-integer laws are continuously interpolated is yet to be clarified.
The finite-size correction 
will be also crucial for  quantum turbulences with 
a large number of vortices, 
since 
 the mean intervortex distance gives 
 the effective system size 
for each vortex.  \\
%
\indent In this paper, we report analytical formulas interpolating 
the integer and non-integer dispersions in finite- and infinite-size systems
for Kelvin and ripple modes, 
and find a complete agreement with numerical simulations. 
We also summarize common practical procedures 
in derivation of these two examples, which could become a guiding principle 
to derive interpolating formulas 
for NGMs around other topological defects.  \\
%
%
	\indent The organization of this paper is as follows. In Sec.~\ref{sec:mainresult}, we summarize our main analytical formulas and their numerical verifications for Kelvin modes and ripplons. We also summarize common aspects of mathematical derivations given in subsequent sections. In Secs.~\ref{sec:dervkelvin} and \ref{sec:ripplon}, we provide full analytical derivations of main results for Kelvin modes and ripplons, respectively. Section~\ref{sec:summary} is devoted to a summary. Appendices~\ref{app:miscripplon} and \ref{app:evaluatemuderiv} provide a few technical calculations for DWs in two-component BECs.
\section{Main Result and Numerical Evidence}\label{sec:mainresult}
\subsection{Kelvin modes}\label{sec:mainresultkel}
\begin{figure}[tb]
		\centering
		\includegraphics[width=0.9\linewidth]{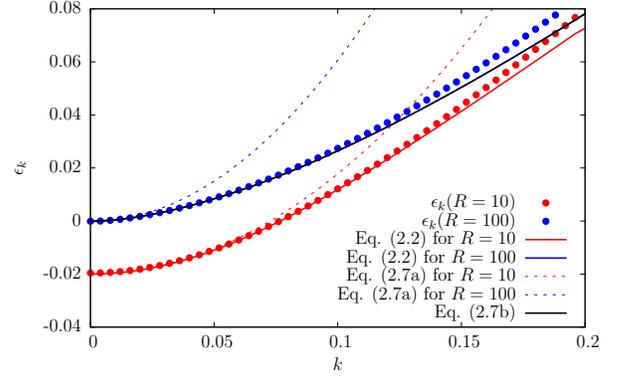}
		\caption{\label{fig:kelvin-dispersion} (Color online) The dispersion $\epsilon_k$ of Kelvin modes for $R=10$ and $R=100$ under the Neumann BC. The analytical formula [Eq. \eqref{eq:Kelvininfinite}] and numerical solutions agree well in the low $k$ region. Equations \eqref{eq:Kelvininfinite} with $R = 100$ and Eq. \eqref{eq:Kelvincasesb} for $ R\to \infty $ are almost the same and the two lines for them in the figure overlap each other.}
	\end{figure}

\begin{figure}[tb]
		\centering
		\includegraphics[width=0.9\linewidth]{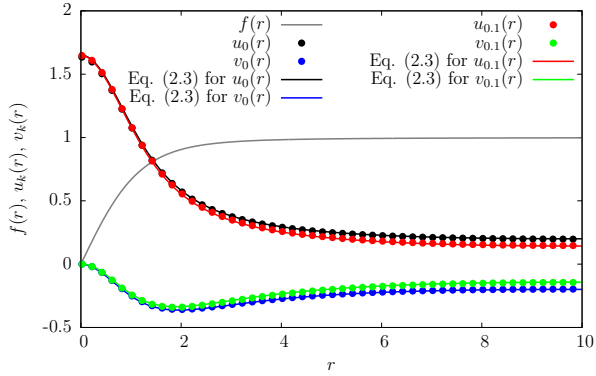}
		\caption{\label{fig:kelzero} (Color online) The zero- and finite-wavenumber ($k = 0.1$) solutions of Kelvin modes under the Neumann BC. Here we set $R=10$.}
	\end{figure}

\begin{figure}[b]
		\centering
		\includegraphics[width=0.9\linewidth]{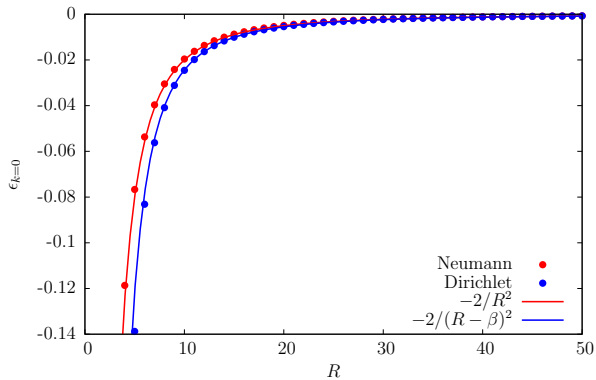}
		\caption{\label{fig:kelvin-epsilonzero}(Color online) The energy shift of the zero-mode solution $ \epsilon_0 $. For the Neumann BC, it is well explained by the direct formula  $ \epsilon_0=-\frac{2}{R^2} $ [Eq.~(\ref{eq:Kelvincasesa})]. For the Dirichlet BC, we find a good fitting if we introduce the ``effective system radius'' $ R-\beta $, where the fitting parameter $ \beta $ is determined to be $ \beta=0.946 $.}
	\end{figure}

\begin{figure}[tb]
		\centering
		\includegraphics[width=0.9\linewidth]{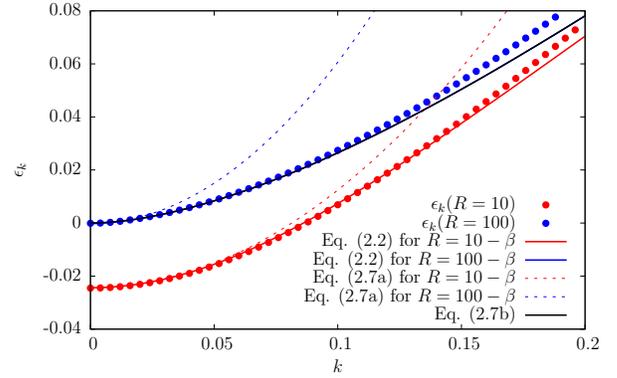}
		\caption{\label{fig:kelvin-dispersion-dirichlet}(Color online) The dispersion relation $\epsilon_k$ of Kelvin modes for $R=10$ and $R=100$ under the Dirichlet BC. Here, when we plot the analytical formulas Eqs. (\ref{eq:Kelvininfinite}) and (\ref{eq:Kelvincasesa}), we use the modified system radius $ R-\beta $ instead of the bare $ R $.}
	\end{figure}

\begin{figure}[tb]
		\centering
		\includegraphics[width=0.9\linewidth]{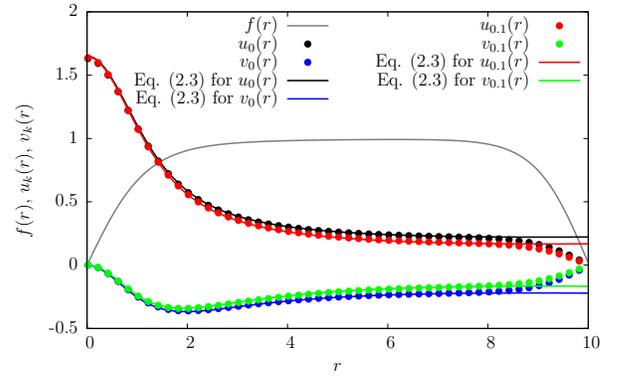}
		\caption{\label{fig:kelzerodirichlet}(Color online) The zero- and finite-wavenumber ($k = 0.1$) solutions of Kelvin modes under the Dirichlet BC with the system radius $ R=10 $. The analytical formula [Eq.~(\ref{eq:kelFr})] is used with replacing $ R $ by $ R-\beta $, where $ \beta\simeq 0.946 $ is obtained from the fitting in Fig.~\ref{fig:kelvin-epsilonzero}. Numerical solutions almost overlap with the analytical solution except near the boundary $ r\simeq R $.}
	\end{figure}

	First we report the interpolating dispersion formula for Kelvin modes propagating along a quantized vortex. The detailed derivations are given in Sec. \ref{sec:dervkelvin}. We consider an infinitely long cylinder with radius $ R $. The Gross-Pitaevskii (GP) energy functional for 
a single-component BEC with chemical potential term is given by
	\begin{align}
		H-\mu N=\int\mathrm{d}^3\boldsymbol{r} \left(\frac{|\nabla\psi|^2}{2m}+g |\psi|^4-\mu |\psi|^2\right).
	\end{align}
	Without loss of generality we set $ 2m=\frac{\mu}{2}=g=1 $ by rescaling of variables. The GP equation is then given by $ \mathrm{i}\partial_t\psi= -\nabla^2\psi-2\psi+2|\psi|^2\psi $. 
	The boundary condition (BC) at $ r=R $ does not affect the main results shown below. For example, it can be either Dirichlet or Neumann. We are interested in a stationary single vortex solution. 
Setting $ \psi=f(r)\mathrm{e}^{\mathrm{i}\theta} $, the function $ f $ satisfies $ -f''-\frac{f'}{r}+\frac{f}{r^2}-2f(1-f^2)=0 $. 
	Henceforth, we write the vortex solution in the infinite-size system ($R=\infty$) as $ f_\infty(r) $. The asymptotic form for large $ r $ is given by $ f_\infty(r)=1-\frac{1}{4r^2}+O(r^{-4}) $. The Bogoliubov equation \cite{Bogoliubov,Fetter197267,DalfovoGiorginiPitaevskiiStringari,PethickSmith} describing quasiparticle excitations is obtained by substituting $ \psi=\psi+u\mathrm{e}^{-\mathrm{i}\epsilon t}+v^*\mathrm{e}^{\mathrm{i}\epsilon^*t} $ into the GP equation and linearizing it with respect to $ (u,v) $. Then, our main result for Kelvin modes is summarized as follows.  The dispersion relation $ \epsilon_k $ and the quasiparticle wavefunctions for Kelvin modes, which we write $ (u,v)=(u_k(r),v_k(r)\mathrm{e}^{-2\mathrm{i}\theta})\mathrm{e}^{\mathrm{i}k_zz} $, are given by 
	\begin{align}
		\epsilon_k=&\,k^2\left( -\log\tfrac{k}{2}+\eta-\gamma-\chi(kR) \right), \label{eq:Kelvininfinite}\\
		\begin{pmatrix} u_k(r) \\ v_k(r) \end{pmatrix}=& \begin{pmatrix} F_k(r)-\frac{1}{r}+\frac{f_\infty(r)}{r}+f_\infty'(r) \\ -F_k(r)+\frac{1}{r}-\frac{f_\infty(r)}{r}+f_\infty'(r) \end{pmatrix}, \label{eq:kelFr}\\
		F_k(r):=&\,k[K_1(kr)+\chi(kR)I_1(kr)],\quad \chi(k):=\tfrac{K_0(k)+K_2(k)}{I_0(k)+I_2(k)}, \label{eq:kelFr2}
	\end{align}
	where $ k=|k_z| $,\  $ I_n, K_n $ are the modified Bessel function of the first and second kind, $ \gamma=0.577\dots $ is the Euler-Mascheroni constant, and $ \eta $ is a constant defined by
	\begin{align}
		\eta:=\int_0^\infty\mathrm{d}r\left[ r f_\infty'(r)^2-2f_\infty(r)f_\infty'(r)\log r \right]\simeq 0.227.
	\end{align}
	Since $ \chi(k) $ has the expansion
	\begin{align}
		\chi(k)=\begin{cases} \frac{2}{k^2}+(-\gamma-\frac{5}{4}-\log\frac{k}{2})+O(k^2) & (k\ll 1) \\ \pi \mathrm{e}^{-2k}[1+\tfrac{7}{4k}+O(k^{-2})] & (k \gg 1), \end{cases} \label{eq:kelbeta}
	\end{align}
	the dispersion formula [Eq.~(\ref{eq:Kelvininfinite})] includes the following two important limiting cases:
	\begin{subnumcases}
		{\epsilon_k \simeq} 
		\!\!\!-\tfrac{2}{R^2}+k^2(\log R+\tfrac{5}{4}+\eta) & $(kR\ll 1)$ \label{eq:Kelvincasesa} \\ 
		\!\!\!k^2(-\log\tfrac{k}{2}+\eta-\gamma) & $(R \rightarrow \infty)$.  \label{eq:Kelvincasesb}
	\end{subnumcases}
	The expression (\ref{eq:Kelvincasesa}) revisits the result of Refs.~\onlinecite{Kobayashi01022014,DTMN}, except for the correction term $ \frac{5}{4}+\eta $ for the $ k^2 $-coefficient. 
	The expression (\ref{eq:Kelvincasesb}) describes the non-integer dispersion $ \epsilon\sim -k^2\log k $ in the infinite volume 
 \cite{Pitaevskii1961,Donnelly}. 
The correction terms including $ \eta $ improve the fitting with numerical results. 
This constant is slightly different from the previously-known value $\frac{1}{4}$ (Ref.~\onlinecite{PhysRev.162.143}); this difference arises from the use of explicit quasiparticle wavefunctions Eq.~(\ref{eq:kelFr}). The equivalent expression for this  $ \eta $ was also reported in Ref.~\onlinecite{Roberts2003}.
	The formula (\ref{eq:Kelvininfinite}) well explains numerical data not only for the above-mentioned limiting cases but also for the intermediate regions. See Fig.~\ref{fig:kelvin-dispersion}. \\ 
	\indent The quasiparticle eigenstate [Eq.~(\ref{eq:kelFr})] with $ R=\infty $ includes Pitaevskii's result \cite{Pitaevskii1961} in two ways; First, setting $ k=0 $, it reduces to $ (u_0(r),v_0(r))=(f_\infty'+\frac{f_\infty}{r},f_\infty'-\frac{f_\infty}{r}) $, which has the physical meaning of the zero-mode solution originated from translational symmetry breaking \cite{DTMN}. (See also Sec. \ref{subsec:kelvinfund} of this paper.) Second, if we focus on the asymptotic region $ r\gg 1 $,  we have $ (u_k(r),v_k(r))\propto K_1(kr) $, which was used to derive $ \epsilon \sim -k^2\log k $ in Ref.~\onlinecite{Pitaevskii1961}. 
	While $ F_k(r) $ has a power series with respect to $ k $ if $ R<\infty $, it becomes invalid for $ R=\infty $, since $ K_1(kr) $ has a logarithmic term. This means that the naive perturbative expansion does not work when $ R=\infty $. 
	Equation~(\ref{eq:kelFr}) well explains the numerical solutions for quasiparticle excitations. See Fig.~\ref{fig:kelzero}. \\ 
	\indent While the numerical results shown in Figs. \ref{fig:kelvin-dispersion} and \ref{fig:kelzero} are those under the Neumann BC [i.e., $f'(R)=u'(R)=v'(R)=0$], our analytical results are also well applicable for the systems obeying the Dirichlet BC [i.e., $f(R)=u(R)=v(R)=0$]. 
	Analytical formulas without any modification can show a modestly good agreement with numerical results even for the Dirichlet BC. 
%
	As we will see below, however, if we introduce an effective system radius $ R-\beta $ with a numerical fitting parameter $ \beta \simeq 0.946 $, 
we obtain a more refined agreement between the numerical results and the analytical formulas.\\
	\indent  Figure~\ref{fig:kelvin-epsilonzero} shows the $ R $-dependence of the energy of zero-wavenumber solution $ \epsilon_0 $. For the Neumann BC, it is well fitted by the formula $ \epsilon_0=-\frac{2}{R^2} $, consistent with Eq.~(\ref{eq:Kelvincasesa}) and Ref.~\onlinecite{Kobayashi01022014}. For the Dirichlet BC, if we fit the numerical result by the ansatz $ -\frac{2}{(R-\beta)^2} $, we find $ \beta\simeq 0.946 $. The physical meaning of this $ \beta $ is obvious; since the Dirichlet BC suppresses the wavefunctions near the boundary, the effective radius 
 gets shorter than that of the Neumann BC 
by a length about the healing length. 
 Figure~\ref{fig:kelvin-dispersion-dirichlet} shows the comparison of dispersion relations between the numerical results and the analytical formulas with $ R $ being replaced by $ R-\beta $. The fitting is improved drastically by using $ R-\beta $ instead of the bare $ R $. Figure~\ref{fig:kelzerodirichlet} shows the quasiparticle wavefunctions, showing a good agreement with the analytical formulas except near the boundary.
	
\subsection{Ripplons}\label{sec:mainresultrip}
\begin{figure}[tb]
		\begin{center}
		\includegraphics[scale=1.05]{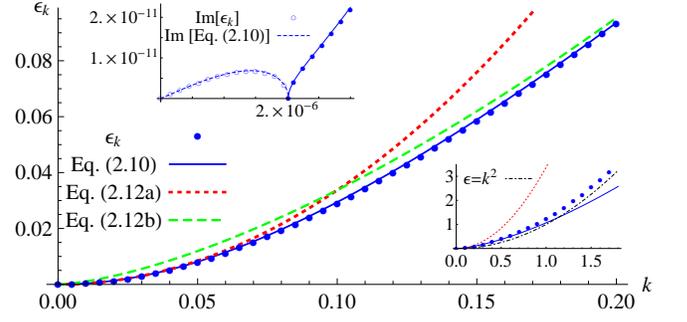}
		\caption{\label{fig:ripneu1}(Color online) Dispersion relations of ripplons in the system under the Neumann BC. We used the following parameters: $2m_1=2m_2=g_{11}=g_{22}=\frac{\mu_1}{2}=\frac{\mu_2}{2}=1,\ g_{12}=2.125,\ $ and $ L=10 $.  $ T_0 $ is numerically calculated as $ 2T_0 \simeq 1.137 $ with assuming $ \partial_d=-\theta(6-|x|)\partial_x $ due to Eq.~(\ref{eq:dderivativemain}).   The upper left inset shows the complex-valued narrow region, and the maximum wavenumber of this region is numerically determined as $ k_c \simeq 2.0\times 10^{-6} $. The lower-right inset shows a plot for larger $ k $'s.}
		\end{center}
	\end{figure}

\begin{figure}[tb]
		\begin{center}
		\includegraphics[scale=.8]{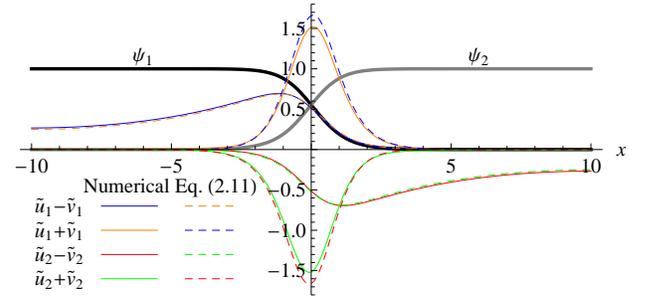}
		\caption{\label{fig:ripneu2}(Color online) Quasiparticle wavefunctions of ripplons in the system under the Neumann BC. The parameters are the same as those of Fig.~\ref{fig:ripneu1}. The wavenumber is $ k=0.2 $. The  $ d $-derivative is approximated by $ \partial_d=-\theta(6-|x|)\partial_x $ due to the relation (\ref{eq:dderivativemain}). }
		\end{center}
	\end{figure}

\begin{figure}[tb]
		\begin{center}
		\includegraphics[scale=.76]{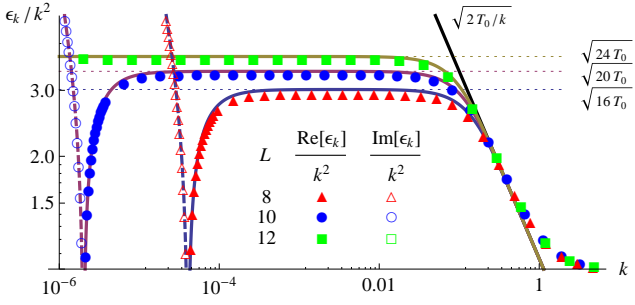} \\
		\caption{\label{fig:ripneu3}(Color online) The log-log plot of  $ k $ vs $ \epsilon_k/k^2 $ under the Neumann BC. The parameters are the same as those of Fig~\ref{fig:ripneu1}. The three solid (dashed) lines show the real (imaginary) part of theoretical formula (\ref{eq:ripploncasesall}) with $ L=8,10,12 $. The plateau region corresponds to quadratic dispersion. The line $ \sqrt{2T_0/k} $ corresponds to the fractional ripplon dispersion [Eq.~(\ref{eq:ripploncasesb})]. $ k_c $'s for $ L=8,10 $ are numerically given by $ k_c\simeq 4.1\times 10^{-5},\ 2.0\times 10^{-6} $, respectively.  $ k_c $ for $ L=12 $ is too small to detect [see Fig.~\ref{fig:ripkc} and the paragraph including Eq. (\ref{eq:kcinmainsec})].} 
		\end{center}
	\end{figure}

\begin{figure}[tb]
		\begin{center}
		\includegraphics[scale=1.05]{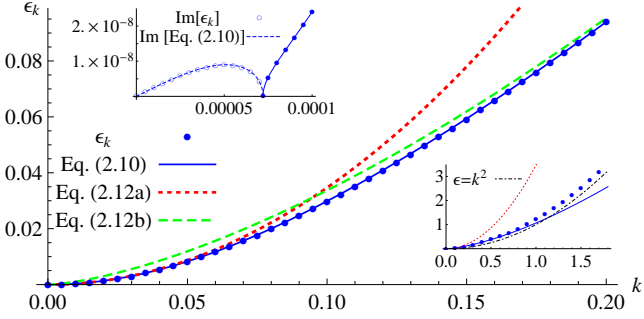}
		\caption{\label{fig:ripdir1}(Color online) The dispersion relation of ripplons in the system under the Dirichlet BC with length $ L=12 $. Here, analytical formulas [Eqs. (\ref{eq:ripploncasesall}), (\ref{eq:ripploncasesa}), and (\ref{eq:ripploncasesb})] are plotted after replacing $ L $ by $ L-\beta\simeq 10.57 $.  $ T_0 $ is numerically calculated as $ 2T_0 \simeq 1.137 $ with assuming $ \partial_d=-\theta(6-|x|)\partial_x $.  The upper left inset shows the complex-valued narrow region. $ k_c\simeq 7.2\times 10^{-5} $ is a numerical fitting parameter. The lower-right inset shows a plot for larger $ k $'s, simply showing that the dispersion relation asymptotically comes close to that of free particles $ \epsilon=k^2 $.}
		\end{center}
	\end{figure}

\begin{figure}[tb]
		\begin{center}
		\includegraphics[scale=.8]{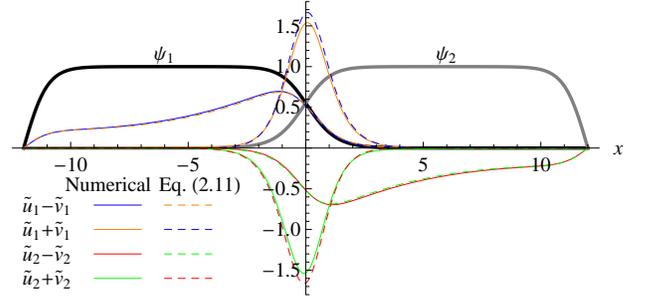} \\
		\caption{\label{fig:ripdir2}(Color online) Quasiparticle wavefunctions of ripplons in the system under the Dirichlet BC with  $ L=12 $. The wavenumber is $ k=0.2 $. In using the theoretical formula [Eq.~(\ref{eq:ripmainuv})], we replace $ L $ by $ L-\beta $.  The  $ d $-derivative is replaced by $ \partial_d=-\theta(6-|x|)\partial_x $. }
		\end{center}
	\end{figure}

\begin{figure}[tb]
		\begin{center}
		\includegraphics[scale=.77]{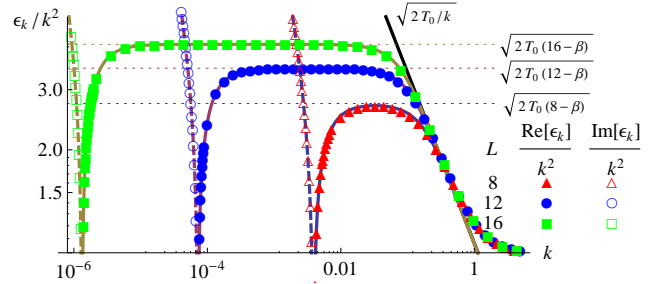} \\
		\caption{\label{fig:ripdir3}(Color online) The log-log plot of $k$ vs $ \epsilon_k/k^2 $ for dispersion relations of ripplons under the Dirichlet BC. The plateau region corresponds to the quadratic dispersion. The line $ \sqrt{2T_0/k} $ corresponds to the fractional ripplon dispersion. The results for $ L=8,12, $ and $ 16 $ are shown. The maximum wavenumbers for the complex region are numerically given by $ k_c=3.9\times 10^{-3},\ 7.2\times 10^{-5},\ 1.3\times 10^{-6} $ for $ L=8,12,16 $, respectively.  Three solid (dashed) lines represent the real (imaginary) part of the analytical formulas [Eq.~(\ref{eq:ripploncasesall})] with $ L $ being replaced by $ L-\beta,\ \beta \simeq 1.43 $. }
		\end{center}
	\end{figure}

\begin{figure}[tb]
		\begin{center}
		\includegraphics[scale=.8]{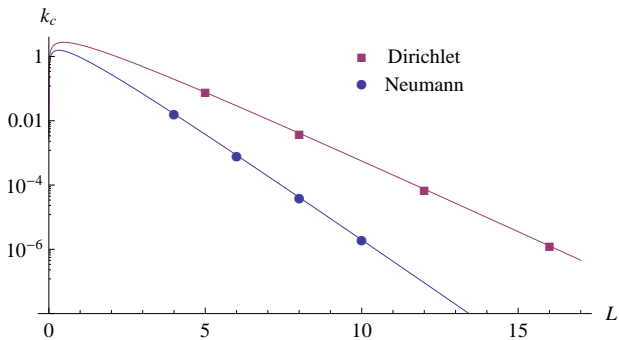}
		\caption{\label{fig:ripkc}(Color online) $ L $-dependence of $ k_c $, the maximum wavenumber of the complex-valued region in the dispersion relation of ripplons. The physical parameters used are the same as other figures. For the Dirichlet BC, the fitting line is given by $ k_c =\sqrt{L}\exp[-1.05L-1.89] $, and that of the Neumann BC is given by $ k_c=\sqrt{L}\exp[-1.58L-1.52] $. }
		\end{center}
	\end{figure}

	Next, we report the dispersion relation of ripplons on a DW 
in two-component BECs. The details of the derivation are given in Sec.~\ref{sec:ripplon}. The energy functional is given by
	\begin{align}
		H=\int\mathrm{d}^3\boldsymbol{r}\left[\sum_{i=1,2}\frac{|\nabla\psi_i|^2}{2m_i}+\sum_{i,j=1,2}g_{ij}|\psi_i|^2|\psi_j|^2\right].
	\end{align} 
	Here we assume $ g_{11},g_{22}>0 $ and $ g_{12}=g_{21}>\sqrt{g_{11}g_{22}} $, in which case the ground state is given by the state such that  $ \psi_1 $ and $ \psi_2 $ are separated \cite{PhysRevA.58.4836,PethickSmith}. We consider the system confined in a cuboid $ [-L_x,L_x]\times[-L_y,L_y]\times[-L_z,L_z] $, and we set a DW perpendicular to the $ x $-axis. Henceforth we simply write $ L_x=L $. The BC can be either Dirichlet or Neumann. 
Mostly we consider the problem with $ L_y=L_z=\infty $. As shown in Sec.~\ref{sec:ripplon}, strictly speaking, the system with $ L_y=L_z=\infty $ has unstable modes, i.e., the Bogoliubov equation has the complex eigenvalue. 
	This instability merely reflects the fact that the true ground states are the states such that the DW is set parallel to the $ x $-axis, because the surface energy becomes smaller for such a configuration. 
	The wavenumbers of unstable modes are, however,  exponentially small $ k_c \sim \mathrm{e}^{-\alpha L} $, and hence we can easily suppress these unstable modes by modifying $ L_y,L_z $ to be very large but finite sizes satisfying $ L \ll L_{y,z} \lesssim \frac{\pi}{k_c} $, 
	which makes the wavenumbers of eigenstates discretized and erases the unstable modes. \\
	\indent Let $ x=d  $ be the position of the DW. By definition $ |d|\le L $ holds.  Let us assume that $ \psi_{1(2)} $ occupies the left (right) side of the DW, and let $ \rho_{1(2)} $ be their densities in the uniform region far from both the boundary and the DW. That means, if we ignore the detailed profiles near the boundary and the DW, the order parameters can be written as $ \psi_1\sim\sqrt{\rho_1}\theta(d-x) $ and $ \psi_2\sim\sqrt{\rho_2}\theta(x-d) $. 
	When $ L $ is large, varying $ d $ with fixed $ \rho_i $'s corresponds to the smooth sliding of the position of DW without changing the profiles of $ \psi_1,\psi_2 $ far from the DW. Therefore, the differentiation of $ \psi_i $'s with respect to $ d $ with fixed $ \rho_i $'s can be approximated as 
	\begin{align}
		\partial_d \simeq \begin{cases} -\partial_x & (x\simeq d) \\ 0 & (|x-d|\gg\xi), \end{cases} \label{eq:dderivativemain}
	\end{align}
	with the typical healing length $ \xi $. In particular, if we take the limit $ L\rightarrow\infty $, we obtain $ \partial_d \rightarrow -\partial_x $. \\ 
	\indent The GP equation is given by $ \mathrm{i}\partial_t\psi_i=\frac{\delta (H-\mu_1N_1-\mu_2N_2)}{\delta \psi_i^*}=\big(-\mu_i-\tfrac{\nabla^2}{2m_i}+2\sum_{j=1,2}g_{ij}|\psi_j|^2\big)\psi_i,\ i=1,2 $. 
	If  $ L $ is large, the values of $ \mu_i $'s are close to those in the infinite-size system:  $ \mu_i \simeq 2g_{ii}\rho_i $. The Bogoliubov equation can be obtained by substituting $ \psi_i=\psi_i+u_i\mathrm{e}^{-\mathrm{i}\epsilon t}+v_i^*\mathrm{e}^{\mathrm{i}\epsilon^*t} $ to the GP equation and linearizing it for $ (u_i,v_i) $. \\
	\indent Now we give our main result on the dispersion relations of ripplons in finite-size systems. 
	For simplicity, here we only present the result for the case $ d=0 $. 
	The general expressions for $ d\ne 0 $ are available in Sec.~\ref{subsec:rippinterpol} [Eqs. (\ref{eq:ripdispmod}), (\ref{eq:rippeigengen}) with (\ref{eq:ripak})]. Let us write the quasiparticle wavefunction as $ (u_1,u_2,v_1,v_2)=(\tilde{u}_1(x),\tilde{u}_2(x),\tilde{v}_1(x),\tilde{v}_2(x))\mathrm{e}^{\mathrm{i}(k_yy+k_zz)} $ and define $ k=(k_y^2+k_z^2)^{1/2} $. Then, the dispersion relation $ \epsilon_k $ and the wavefunction of the ripplon are given by
	\begin{align}
		\epsilon_k&=\sqrt{\frac{2T_0}{m_1\rho_1+m_2\rho_2}\frac{\tanh kL}{k}k^2(k^2-k_c^2)}, \label{eq:ripploncasesall}\\
		\begin{pmatrix}\tilde{u}_1 \\ \tilde{u}_2 \\ \tilde{v}_1 \\ \tilde{v}_2 \end{pmatrix}&=\frac{\epsilon_k}{k\cosh kL}\begin{pmatrix}m_1\cosh k(x+L)\psi_1 \\ -m_2\cosh k(x-L)\psi_2 \\ -m_1\cosh k(x+L)\psi_1^* \\ m_2\cosh k(x-L)\psi_2^* \end{pmatrix}+\tanh kL \begin{pmatrix}\partial_d\psi_1 \\ \partial_d\psi_2 \\ \partial_d\psi_1^* \\ \partial_d\psi_2^* \end{pmatrix}, \label{eq:ripmainuv}
	\end{align}
	where $ k_c\sim O(\mathrm{e}^{-\alpha L}) $ is the maximum wavenumber of unstable modes mentioned above, and $ T_0=\int\mathrm{d}x\Bigl( \frac{|\partial_d\psi_1|^2}{2m_1}+\frac{|\partial_d\psi_2|^2}{2m_2} \Bigr) $ represents the tension of the DW, recalling the relation Eq.~(\ref{eq:dderivativemain}). If we ignore the narrow complex region $ k \le k_c $, the dispersion relation includes the following two cases:
	\begin{subnumcases}
		{\epsilon_k \simeq \sqrt{\frac{2T_0}{m_1\rho_1+m_2\rho_2}}\times } 
		\!\!\!\sqrt{L}k^2 & $(kL \ll 1)$ \label{eq:ripploncasesa} \\ 
		\!\!\!k^{3/2} & $(L\rightarrow\infty)$.  \label{eq:ripploncasesb}
	\end{subnumcases}
	The behavior $ \epsilon \sim \sqrt{L}k^2 $ is consistent with Refs.~\onlinecite{PhysRevA.88.043612,DTMN}, and the latter case (\ref{eq:ripploncasesb}) describes the fractional dispersion relation \cite{PhysRevA.65.033618,PhysRevA.88.043612}. The quasiparticle eigenfunction Eq.~(\ref{eq:ripmainuv}) in the limit $ L\rightarrow\infty $ is given by
	\begin{align}
		\begin{pmatrix}\tilde{u}_1 \\ \tilde{u}_2 \\ \tilde{v}_1 \\ \tilde{v}_2 \end{pmatrix} = -\begin{pmatrix}\partial_x\psi_1 \\ \partial_x\psi_2 \\ \partial_x\psi_1^* \\ \partial_x\psi_2^*  \end{pmatrix}+\sqrt{\frac{2T_0k}{m_1\rho_1+m_2\rho_2}} \begin{pmatrix} m_1\psi_1\mathrm{e}^{kx} \\ -m_2\psi_2\mathrm{e}^{-kx} \\ -m_1\psi_1^*\mathrm{e}^{kx} \\ m_2\psi_2^*\mathrm{e}^{-kx} \end{pmatrix}.
	\end{align}
	with recalling $ \partial_d \rightarrow -\partial_x $ [Eq.~(\ref{eq:dderivativemain})]. It describes the quasiparticle wavefunction of ripplons in the infinite system. The former term is the zero-mode solution originated from translational symmetry breaking. The latter term represents the oscillation of relative phases between $ \psi_1 $ and $ \psi_2 $ and includes $ \sqrt{k} $, indicating that the naive perturbation is impossible.\\ 
	\indent Let us see the numerical evidence for the above analytical results. We first show the result for the Neumann BC. Figure~\ref{fig:ripneu1} shows the numerical verification of dispersion relations. An example of quasiparticle wavefunctions is given in Fig.~\ref{fig:ripneu2}. The $ L $-dependence of the quadratic and complex dispersion regions is well illustrated by plotting the $ k $-dependence of $ \epsilon_k/k^2 $. See Fig.~\ref{fig:ripneu3}. \\ 
%
%
		\indent Our analytical formulas also explain the numerical results for the Dirichlet BC. As with the case of Kelvin modes, we find that the replacement of the effective system length $ L \rightarrow L-\beta $ with $ \beta\simeq 1.43 $, and this replacement is used in plotting the analytical formulas. Figure \ref{fig:ripdir1} shows the comparison of dispersion relations between numerical data and analytical formulas with $ L $ being replaced by $ L-\beta $. Even when we use the bare $ L $, a modestly good agreement with the numerical data is obtained. However, if we use the modified $ L-\beta $, the fitting becomes rather perfect. Figure~\ref{fig:ripdir2} shows the wavefunctions of quasiparticle eigenstates. Figure~\ref{fig:ripdir3} shows the log-log plot of $ \epsilon_k/k^2 $, in which the  $ L $-dependence of the quadratic dispersion relation becomes visible. The value of $ \beta $ is evaluated from the plateau region of the data of $ L=12 $ and $ 16 $ in this figure. \\
	\indent Here, we give a few additional remarks on the width of the complex-valued regions in the dispersion relation, i.e., $ k_c $ in Eq.~(\ref{eq:ripploncasesall}). As derived in Appendix~\ref{app:evaluatemuderiv}, if we consider the system such that $ 2m_1=2m_2=g_{11}=g_{22}=1 $,\ $ g_{12}=\infty $, and the average density is given by $ \rho_0=1 $, the $ L $-dependencies of $ k_c $ for the Dirichlet and the Neumann BCs are given by
	\begin{align}
		k_c \propto \begin{cases} \sqrt{L}\mathrm{e}^{-L} & \text{(Dirichlet)}, \\ \sqrt{L}\mathrm{e}^{-2L}  & \text{(Neumann)}. \end{cases} \label{eq:kcinmainsec}
	\end{align}
	Thus, $ k_c $ in the systems under the Neumann BC decreases more rapidly than that under the Dirichlet BC. This relation can be also confirmed for finite $ g_{12} $ with a slight modification of the coefficients in exponential factors.  See Fig.~\ref{fig:ripkc}. From this figure, we can understand why we cannot find $ k_c $ in the system with the Neumann BC with length $ L=12 $ in Fig.~\ref{fig:ripneu3}. We expect $ k_c \simeq 9\times 10^{-8} $ from Fig.~\ref{fig:ripkc}, implying that the typical eigenenergy of complex-valued region is  $ |\epsilon| \sim O(k_c^2)\sim O(10^{-15}) $. This is too small to determine $ k_c $ precisely in the double-precision calculation. These results are consistent with Ref.~\onlinecite{PhysRevA.88.043612}, where the numerical simulations with very large $ L $'s were performed under the Neumann BC, and complex eigenvalues were not found. 
\subsection{Sketch of derivation: summarizing common procedures}\label{subsec:sketch}
	Having presented our main results, we briefly summarize common procedures of detailed derivations, which will be given in Secs. \ref{sec:dervkelvin} and \ref{sec:ripplon}. 
	Even though the mathematical justifications for each example are slightly different, the practical procedures are similar. 
	They are summarized as follows:
	\begin{enumerate}[(A)]
		\item First, derive zero-mode solutions 
having the origin of spontaneous symmetry breaking (SSB) 
in the infinite system \cite{DTMN}.
		\item In the intermediate region far from both topological defects and the boundary, where the asymptotic form of the order parameter becomes almost exact, derive the finite-wavenumber solution of the Bogoliubov equation. In such a region where the local structure of the order parameter is ignorable, the density fluctuation ($ \sim u+v $) becomes irrelevant compared to the phase fluctuation ($ \sim u-v $), and hence the differential equation becomes solvable. Here, the integration constants are fixed by assuming the Neumann BC $ \lim_{\boldsymbol{r}\rightarrow \text{boundary}}\boldsymbol{n}\cdot \nabla u(\boldsymbol{r})=\boldsymbol{n}\cdot \nabla v(\boldsymbol{r})=0 $. 
		\item Make a minimal modification to the solution obtained in (B) to include the exact zero-mode solutions derived in (A)  to take into account the local structure near the topological defects.
		\item Using the solution constructed in the above way, calculate an eigenenergy $ \epsilon_k $ solving the Bogoliubov equation by using the techniques in Ref.~\onlinecite{DTMN}.
	\end{enumerate}
	Here, we emphasize that the use of the Neumann BC in the procedure (B) does \textit{not} mean that our result is not applicable for other BCs, e.g., the Dirichlet BC. The purpose of (B) is to obtain the quasiparticle wavefunctions in the asymptotic region where the behavior of the order parameter becomes almost uniform. 
	Since $ u,v $ are linearized fields of the order parameter, they also should obey the same uniform boundary condition, and hence the Neumann BC is most suitable for this purpose. 
	To be more concrete, let  $ |\boldsymbol{r}| $  be a distance from a topological defect and let $ \xi $ and $ L $ be a typical healing length and the distance between the defect and the boundary, respectively.  Then, the solution obtained in (B) is quite applicable in the intermediate region $ \xi \ll |\boldsymbol{r}| \lesssim L-\xi $, and the behavior in this region is independent of the choice of BCs. The behaviors of quasiparticle wavefunctions $ (u(\boldsymbol{r}),v(\boldsymbol{r})) $ very near the boundary $ L-\xi \lesssim |\boldsymbol{r}| \le L $ gives no influence to the leading order of the dispersion relation $ \epsilon_k $ and is not of our interest in the current problem. By the modification (C), the solution becomes applicable even near a topological defect, i.e., $ 0 \le |\boldsymbol{r}| \lesssim L-\xi $, and thus the effects of zero modes are correctly included. For the example of the Kelvin modes, the procedure (B) gives the solution $ F_k(r) $ [Eq.~(\ref{eq:kelFr2})], and the procedure (C) gives Eq.~(\ref{eq:kelFr}). For the ripplons, (B) gives $ \cosh k(x\pm L) $ and (C) gives Eq.~(\ref{eq:ripmainuv}). See Secs. \ref{sec:dervkelvin} and \ref{sec:ripplon} for detailed derivations. The evidence of applicability for both Neumann and Dirichlet BCs is actually presented in the former part of this section. 
	
	Note that, if we consider NGMs concerning spin degree of freedom, the terms ``density fluctuation'' and ``phase fluctuation'' in the procedure (B) should be replaced by ``fluctuation of the magnitude of magnetization'' and ``fluctuation of the angle of magnetization,'' respectively.

\section{Detailed derivation --- Kelvin modes}\label{sec:dervkelvin}
	Thus far, we have presented our main analytical formulas and their numerical verifications for Kelvin and ripple modes. In this and next section, we provide the complete derivations of these formulas.
\subsection{Fundamental equations and zero modes}\label{subsec:kelvinfund}
	The energy functional of the one-component BEC in the dimensionless form is given by
	\begin{align}
		H-\mu N=\int\mathrm{d}^3\boldsymbol{r}\left( |\nabla\psi|^2+|\psi|^4-2|\psi|^2 \right).
	\end{align}
	The stationary GP and Bogoliubov equations are 
	\begin{align}
		(-\nabla^2-2+2|\psi|^2)\psi&=0,\\
		\begin{pmatrix} -\nabla^2-2+4|\psi|^2 & 2\psi^2 \\ -2\psi^{*2} & \nabla^2+2-4|\psi|^2 \end{pmatrix}\begin{pmatrix}u \\ v\end{pmatrix}&=\epsilon\begin{pmatrix}u \\ v\end{pmatrix}.
	\end{align}
	Since we are interested in the vortex solution with the vortex charge $ n=1 $, we set $ \psi=f(r)\mathrm{e}^{\mathrm{i}\theta} $, where $ f(r) $ is a non-negative function having the asymptotic form $ f(\infty)=1 $. Then the GP equation becomes
	\begin{align}
		-f''-\frac{f'}{r}+\frac{f}{r^2}-2f(1-f^2)=0. \label{eq:supkelgp1}
	\end{align}
	Henceforth we write the solution in the infinite-size system as $ f_\infty(r) $. The asymptotic solution is given by
	\begin{align}
		f_\infty(r)=1-\frac{1}{4r^2}-\frac{9}{32r^4}+O(r^{-6}). \label{eq:supkelgp2}
	\end{align}
	The expansion at $ r=0 $ can be also obtained, given by
	\begin{align}
		f_\infty(r)=ar-\frac{a}{4}r^3+\frac{a+4a^3}{48}r^5+O(r^7), \label{eq:supkelgp25}
	\end{align}
	where $ a \simeq 0.82 $ is a constant determined numerically.\\ 
	\indent In the infinite-size system, the GP equation has a symmetry such that ``$ \psi(x,y,z) $ is a solution'' $ \leftrightarrow $ ``$ \psi(x+x_0,y+y_0,z)\mathrm{e}^{\mathrm{i}\theta} $ is also a solution''. Differentiating the GP equation by $ \theta,x_0, $ and $ y_0 $, we obtain the following SSB-originated zero mode solutions \cite{DTMN} for the Bogoliubov equation:
	\begin{align}
		w_{\text{phase}}=\begin{pmatrix}\psi \\ -\psi^* \end{pmatrix},\ w_{x\text{-trans}}=\begin{pmatrix}\partial_x\psi \\ \partial_x\psi^*\end{pmatrix},\ w_{y\text{-trans}}=\begin{pmatrix}\partial_y\psi \\ \partial_y\psi^*\end{pmatrix}.
	\end{align}
	As shown in Ref.~\onlinecite{DTMN},  $ w_{\text{phase}} $ is  $ \sigma $-orthogonal to the other two zero modes, so it solely yields a type-I NGM, which is the Bogoliubov phonon. On the other hand,  $ w_{x\text{-trans}} $ and $ w_{y\text{-trans}} $ are not $ \sigma $-orthogonal and becoming a pair yielding one type-II NGM, the Kelvin mode. We can construct a positive-norm zero-mode solution becoming a seed of type-II mode by their linear combination, which is given by\cite{DTMN}
	\begin{align}
		w_0&=w_{x\text{-trans}}-\mathrm{i}w_{y\text{-trans}}=\begin{pmatrix} f_\infty'+\frac{f_\infty}{r} \\ \mathrm{e}^{-2\mathrm{i}\theta}(f_\infty'-\frac{f_\infty}{r}) \end{pmatrix}. \label{eq:supkelw0}
	\end{align}
	Then $ w_0 $ becomes the seed of the positive dispersion branch. The same solution was also derived by Pitaevskii \cite{Pitaevskii1961}. $ w_{x\text{-trans}}+\mathrm{i}w_{y\text{-trans}} $  has negative norm and yields the negative dispersion branch. \\ 
	\indent The Bogoliubov equation can be decoupled for different angular momenta by setting $ (u,v)=(u(r,z)\mathrm{e}^{\mathrm{i}\theta},v(r,z)\mathrm{e}^{-\mathrm{i}\theta})\mathrm{e}^{\mathrm{i}m\theta}, $ \ $ m \in \mathbb{Z} $.  We are further interested in the solution propagating in the $ z $-direction. So we set $ (u(r,z),v(r,z))=\mathrm{e}^{\mathrm{i}k_zz}(u(r),v(r)) $. The resultant equation is
	\begin{align}
		&\epsilon\begin{pmatrix}u \\ v \end{pmatrix}=(H_0+\sigma k^2)\begin{pmatrix}u \\ v \end{pmatrix} \label{eq:kelfundH00} 
	\end{align}
	with $ k=|k_z| $,  $ \sigma=\operatorname{diag}(1,-1) $ and
	\begin{align}
		[H_0]_{11}&=-\partial_r^2-\frac{1}{r}\partial_r+\frac{(m+1)^2}{r^2}-2+4f^2, \label{eq:kelfundH11} \\
		[H_0]_{12}&=-[H_0]_{21}=2f^2,\\
		[H_0]_{22}&=\partial_r^2+\frac{1}{r}\partial_r-\frac{(m-1)^2}{r^2}+2-4f^2. \label{eq:kelfundH22}
	\end{align}
	The Kelvin mode with positive dispersion exists in the sector $ m=-1 $, since it contains the zero-mode $ w_0 $ [Eq.~(\ref{eq:supkelw0})]. Henceforth we consider only this sector. The asymptotic behavior of zero-mode solution is given by
	\begin{align}
		u&=f_\infty'(r)+\frac{f_\infty(r)}{r}=\frac{1}{r}+\frac{1}{4r^3}+\dotsb, \label{eq:kelzero11} \\
		v&=f_\infty'(r)-\frac{f_\infty(r)}{r}=-\frac{1}{r}+\frac{3}{4r^3}+\dotsb. \label{eq:kelzero12}
	\end{align}
	The  $ \sigma $-inner products between two quasiparticle wavefunctions $ w_i=(u_i(r),v_i(r)),\ i=1,2 $ is defined by
	\begin{align}
		(w_1,w_2)_\sigma:=\int_0^Rr\mathrm{d}r(u_1^*u_2-v_1^*v_2).
	\end{align}
	Here we omit the $ \theta $-integration, which merely gives the factor $ 2\pi $ in this problem. $ H_0 $ satisfy the following property
	\begin{align}
		(w_1,H_0w_2)_\sigma=(H_0w_1,w_2)_\sigma, \label{eq:kelvh0bhr}
	\end{align}
	which holds for any ``Bogoliubov-hermitian'' operator \cite{DTMN}, and can be regarded as an analog of self-adjointness for hermitian operators. Using these inner products and analog of self-adjointness, we can construct a perturbation theory in a similar way to that of ordinary hermitian operators \cite{DTMN}. 
\subsection{Type-II dispersion coefficient in finite systems}
	Henceforth we consider the finite-size systems. Let the system be an infinitely-long cylinder with finite radius $ R $. The BC at $ r=R $ is arbitrary and does not give an influence to the following argument. In a finite-size system, the translational symmetry no longer exists and hence $ w_{x\text{-trans}} $ and $ w_{y\text{-trans}} $ do not become the exact zero-mode solutions. Let us see how these zero-mode solutions are modified in finite-size systems. \\
	\indent We solve the Bogoliubov equation using the expansion w.r.t the parameter $ \alpha:=R^{-1} $. Then $ \alpha=0 $ corresponds to the infinite-size system $ R=\infty $ and finite $ \alpha $ corresponds to finite-size systems. Let us write $ \xi=r/R=r\alpha $. Then  $ \xi $ can take a value in the closed interval $ [0,1] $. Let us further write $ \tilde{f}(\xi,\alpha):=f(\xi/\alpha) $. We henceforth use the prime symbol to express the  $ \xi $-derivative, e.g.,  $ \tilde{f}'=\frac{\mathrm{d}\tilde{f}}{\mathrm{d}\xi} $. Then the GP equation (\ref{eq:supkelgp1}) becomes
	\begin{align}
		\alpha^2\left( -\tilde{f}''-\frac{\tilde{f}'}{\xi}+\frac{\tilde{f}}{\xi^2} \right)-2\tilde{f}(1-\tilde{f}^2)=0.
	\end{align}
	Let us seek a solution in the form of $ \alpha $-expansion:  $ \tilde{f}=\tilde{f}_0+\alpha^2\tilde{f}_2+\alpha^4\tilde{f}_4+\dotsb $. Note that the expansion around $ \alpha=0 $ is rather sensitive and only meaningful in $ 0<\xi<1 $. At $ \xi=0 $ and $ 1 $, the expansion is pathological and we do not consider it. Here, we are only interested in the intermediate regions far from both the vortex and the boundary. The GP equations for each order then become
	\begin{align}
		\alpha^0:\quad& \tilde{f}_0(1-\tilde{f}_0^2)=0,\\
		\alpha^2:\quad& -\tilde{f}_0''-\frac{\tilde{f}_0'}{\xi}+\frac{\tilde{f}_0}{\xi^2}-2\tilde{f}_2(1-3\tilde{f}_0^2)=0, \\
		\alpha^4:\quad& -\tilde{f}_2''-\frac{\tilde{f}_2'}{\xi}+\frac{\tilde{f}_2}{\xi^2}-2\tilde{f}_4(1-3\tilde{f}_0^2)+6\tilde{f}_0\tilde{f}_2^2=0. 
	\end{align}
	The solution satisfying the asymptotic condition $ f(r\rightarrow\infty)=1 $ is given by $ \tilde{f}_0=1 $, and  $ \tilde{f}_2,\tilde{f}_4,\dots $ are determined iteratively:
	\begin{align}
		\tilde{f}_0=1,\quad \tilde{f}_2=-\frac{1}{4\xi^2},\quad \tilde{f}_4=-\frac{9}{32\xi^4},\quad\dots
	\end{align}
	Thus we have
	\begin{align}
		\tilde{f}=1-\alpha^2\frac{1}{4\xi^2}-\alpha^4\frac{9}{32\xi^4}+O(\alpha^6).
	\end{align}
	This is just the revisit of Eq. (\ref{eq:supkelgp2}). \\ 
	\indent Next we solve the Bogoliubov equation by the same expansion. The Bogoliubov equation rewritten by $ \xi $ and $ \alpha $ is given by
	\begin{align}
		\alpha^2\left( -u''-\frac{u'}{\xi} \right)-2(1-2\tilde{f}^2)u+k^2u+2\tilde{f}^2v&=\epsilon u, \label{eq:albogou} \\
		\alpha^2\left( v''+\frac{v'}{\xi}-\frac{4v}{\xi^2} \right)+2(1-2\tilde{f}^2)v+k^2v-2f^2u&=\epsilon v. \label{eq:albogov}
	\end{align}
	Here we again note that the prime represents the differentiation by $ \xi $. \\
	\indent We first consider the zero-wavenumber case $ k=0 $ and examine the energy shift of the zero-mode solution $ w_0 $ due to the finite-size effect. Let $ \epsilon_0 $ be the energy shift of the zero-mode solution, and let us expand it as $ \epsilon_0=\epsilon_{0,0}+\alpha^2\epsilon_{0,2}+\alpha^4\epsilon_{0,4}+\dotsb $. We already know that the eigenvalue of $ w_0 $ in the infinite system  $ (\alpha=0) $ is zero: $ \epsilon_{0,0}=0 $. We also expand the quasiparticle wavefunctions in the same way: $ u=u_0+\alpha^2 u_2+\dotsb,\ v=v_0+\alpha^2v_2+\dotsb $. The zeroth- and second-order equations are then given by
	\begin{align}
		u_0+v_0&=0, \\
		\quad -u_0''-\frac{u_0'}{\xi}-\frac{u_0}{\xi^2}+2(u_2+v_2)&=\epsilon_{0,2}u_0, \label{eq:kelnaiv01}\\
		\quad v_0''+\frac{v_0'}{\xi}-\frac{3v_0}{\xi^2}-2(u_2+v_2)&=\epsilon_{0,2}v_0. \label{eq:kelnaiv02}
	\end{align}
	Thus we obtain $ u_0+v_0=0 $, which justifies ignoring the density fluctuation in the procedure (B) of Sec.~\ref{subsec:sketch}.  
	Taking the sum and difference of Eqs. (\ref{eq:kelnaiv01}) and (\ref{eq:kelnaiv02}), and using $v_0=-u_0$, we obtain
	\begin{align}
		-u_0''-\frac{u_0'}{\xi}+\frac{u_0}{\xi^2}&=0, \label{eq:kelnaivsum}\\
		-\frac{2u_0}{\xi^2}+2(u_2+v_2)&=\epsilon_{0,2}u_0. \label{eq:kelnaivdiff}
	\end{align}
	The solution of Eq.~(\ref{eq:kelnaivsum}) is given by $ u=c_1\xi+c_2\xi^{-1} $. Following the procedure (B), we fix the coefficient by the Neumann BC $ u_0'(\xi\rightarrow 1)=0 $. Thus,
	\begin{align}
		u_0=-v_0=\xi+\frac{1}{\xi}. \label{eq:kelnaivlead}
	\end{align}
	If we go back to the original variables,  $ r $ and $ R $, this solution can be rewritten as
	\begin{align}
		u_0=-v_0=\frac{1}{r}+\frac{r}{R^2}. \label{eq:kelnaivlead2}
	\end{align}
	While the term $ \frac{1}{r} $ corresponds to the expansion of the zero-mode solution in the infinite-system Eqs.~(\ref{eq:kelzero11}) and (\ref{eq:kelzero12}), the latter term  $ \frac{r}{R^2} $ exists purely by the finite-size effect. This term is necessary to obtain the energy shift $ \epsilon_{0,2} $. \\ 
	\indent Let us find the expansion coefficient $ \epsilon_{0,2} $. To derive this, we focus on  Eq.~(\ref{eq:kelnaivdiff}) in the region  $ 0<\xi \ll 1 $. Using the next leading orders of Eqs.~(\ref{eq:kelzero11}) and (\ref{eq:kelzero12}), in the region $ 0<\xi \ll 1 $, the leading order terms of $ u_2 $ and $ v_2 $ are given by
	\begin{align}
		u_2 = \frac{1}{4\xi^3}+O(\xi^{-1}),\quad v_2=\frac{3}{4\xi^3}+O(\xi^{-1}). \label{eq:kelnaivnext}
	\end{align}
	Substituting Eqs. (\ref{eq:kelnaivlead}) and (\ref{eq:kelnaivnext}) to Eq. (\ref{eq:kelnaiv01}) and comparing the coefficient of  $ \xi^1 $ in both sides, we obtain $ \epsilon_{0,2}=-2 $. Thus, the energy shift of the zero-mode solution in the finite-size system becomes
	\begin{align}
		\epsilon_0 =-\frac{2}{R^2}+O(R^{-4}) \quad \text{(Neumann BC)}, \label{eq:kelepsilonzero}
	\end{align}
	as with Ref.~\onlinecite{Kobayashi01022014}. If we use the Dirichlet BC, we find a little larger correction due to the boundary effect:
	\begin{align}
		\epsilon_0 =-\frac{2}{R^2}+O(R^{-3}) \quad \text{(Dirichlet BC)}, \label{eq:kelepsilonzerodiri}
	\end{align}
	though the leading order is the same. \\
	\indent The solution (\ref{eq:kelnaivlead2}) well describes the numerical solution in the region $ 0\ll r\lesssim R $, but it diverges at $ r=0 $. 
	This artificial divergence is caused by the fact that the $ \alpha $-expansion is valid only for $ \xi\in(0,1) $. 
	In order to get the correct behavior near the vortex core $ r=0 $, we heuristically replace the divergent term $ r^{-1} $ by the zero-mode solution of infinite systems, i.e., Eqs.~(\ref{eq:kelzero11}) and (\ref{eq:kelzero12}). This replacement is good if the system size $ R $ is sufficiently large, because the profile of quasiparticle wavefunctions near the vortex core is almost the same with those of infinite-size systems. Thus, we obtain the modified zero-mode solution in the finite-size system as 
	\begin{align}
		w_0=\begin{pmatrix}u_0 \\ v_0\end{pmatrix}=\frac{1}{\sqrt{2}}\begin{pmatrix}f_\infty'+\frac{f_\infty}{r}+\frac{r}{R^2} \\ f_\infty'-\frac{f_\infty}{r}-\frac{r}{R^2} \end{pmatrix}, \label{eq:kelmodzeromode}
	\end{align}
	where the factor $ \frac{1}{\sqrt{2}} $ is a normalization factor. This modification corresponds to the procedure (C) in Sec.~\ref{subsec:sketch}. \\
	\indent Using Eq.~(\ref{eq:kelmodzeromode}), we can calculate the coefficient of type-II dispersion. Let us solve the Bogoliubov equation perturbatively:
	\begin{align}
		&(H_0+\sigma k^2)(w_0+k^2w_2+\dotsb)\nonumber \\
		&\quad=(\epsilon_0+k^2\epsilon_2+\dotsb)(w_0+k^2w_2+\dotsb).
	\end{align}
	The zeroth and the second order equations are
	\begin{align}
		H_0w_0&=\epsilon_0w_0, \\
		H_0w_2+\sigma w_0&=\epsilon_0w_2+\epsilon_2w_0.
	\end{align}
	Here we already know $ \epsilon_0=-\frac{2}{R^2}+O(R^{-4}) $. Note that $ \epsilon_2 $ in this $ k $-expansion is different from  $ \epsilon_{0,2} $ appearing in the $ \alpha $-expansion of $ \epsilon_0 $. Taking the  $ \sigma $-inner product between $ w_0 $ and the second order equation, and using (\ref{eq:kelvh0bhr}), we have
	\begin{align}
		\epsilon_2=\frac{(w_0,\sigma w_0)_\sigma}{(w_0,w_0)_\sigma}=\frac{\int_0^Rr\mathrm{d}r(u_0^2+v_0^2)}{\int_0^Rr\mathrm{d}r(u_0^2-v_0^2)}.
	\end{align}
	The denominator is evaluated as
	\begin{align}
		(w_0,w_0)_\sigma&=\int_0^R\mathrm{d}r\left[2f_\infty'(r)\left( f_\infty(r)+\frac{r^2}{R^2} \right)\right]\nonumber \\
		&=1+O\left(\frac{\log R}{R^2}\right), \label{eq:w0sigmaw0}
	\end{align}
	where the orders of each term are evaluated using Eq.~(\ref{eq:supkelgp2}):
	\begin{align}
		\int_0^R\mathrm{d}r\left[2f_\infty'(r)f_\infty(r)\right]&=f_\infty(R)^2=1+O(R^{-2}), \\
		\int_0^R\mathrm{d}r\left[f_\infty'(r)r^2\right]&\sim \int^R\mathrm{d}r\left[ \frac{1}{r^3}r^2 \right]\sim \log R.
	\end{align}
	Thus,  $ w_0 $ is normalized up to $ O(R^{-2}\log R) $ terms. The numerator is given by
	\begin{align}
		(w_0,\sigma w_0)_\sigma&=\int_0^R\mathrm{d}r\left[ \frac{r^3}{R^4}+\frac{2rf_\infty(r)}{R^2}+r f_\infty'(r)^2+\frac{f_\infty(r)^2}{r} \right].
	\end{align}
	The leading orders of each term are given by
	\begin{align}
		\int_0^R\mathrm{d}r\left[ \frac{r^3}{R^4} \right]&=\frac{1}{4},\label{eq:kelnaivcoef3} \\
		\int_0^R\mathrm{d}r\left[ \frac{2rf_\infty(r)}{R^2} \right]&=1+O\left( \frac{\log R}{R^2} \right), \label{eq:kelnaivcoef4} \\
		\int_0^R\mathrm{d}r\left[ r f_\infty'(r)^2 \right]&=\int_0^\infty\mathrm{d}r\left[ r f_\infty'(r)^2 \right]+O(R^{-4}), \label{eq:kelnaivcoef1} 
	\end{align}
	and
	\begin{align}
		&\int_0^R\mathrm{d}r\left[ \frac{f_\infty(r)^2}{r} \right]\nonumber \\
		=&\left[f_\infty(r)^2\log r  \right]_0^R-\int_0^R\mathrm{d}r\left[ 2f_\infty(r)f_\infty'(r)\log r \right] \nonumber \\
		=&\log R-\int_0^\infty\mathrm{d}r\left[ 2f_\infty(r)f_\infty'(r)\log r \right]+O\left( \frac{\log R}{R^2} \right).
	\end{align}
	Here, $ f_\infty(r)^2\log r|_{r=0} $ vanishes since $ f_\infty(r) \simeq  ar $ [Eq.~(\ref{eq:supkelgp25})]. 
	Thus, we obtain
	\begin{align}
		&(w_0,\sigma w_0)_\sigma=\log R+\frac{5}{4}+\eta+O\left( \frac{\log R}{R^2} \right), \label{eq:w0sigmsigmaaw0}\\
		&\eta:=\int_0^\infty\mathrm{d}r\left[ r f_\infty'(r)^2-2f_\infty(r)f_\infty'(r)\log r \right]\simeq 0.227.
	\end{align}
	A closed form for this $ \eta $ is not known. \\
	\indent Summarizing, we obtain
	\begin{align}
		\epsilon&=-\frac{2}{R^2}+A k^2+O(k^4) \label{eq:kelzeronaiv301}
	\end{align}
	with
	\begin{align}
		A&=\log R+\frac{5}{4}+\eta+O\left( \frac{\log R}{R^2} \right) \quad \text{(Neumann BC)}. \label{eq:kelzeronaiv302}
	\end{align}
	If we use the Dirichlet BC [$ f(R)=u(R)=v(R)=0 $], the profile of quasiparticle wavefunctions near the boundary $ r\simeq R $ deviates from $ w_0=(u_0,v_0)^T $. This deviation yields a little larger correction:
	\begin{align}
		A=\log R+\frac{5}{4}+\eta+O(R^{-1}) \quad \text{(Dirichlet BC)}. \label{eq:kelzeronaiv301d}
	\end{align}
	In both cases, however, the leading term is the same. \\
	\indent As shown in Eqs. (\ref{eq:kelepsilonzerodiri}) and (\ref{eq:kelzeronaiv301d}), the Dirichlet BC gives a little larger deviation from the leading order term compared to the Neumann BC. As discussed in Sec.~\ref{sec:mainresult}, these deviations are well included by the effective replacement
	\begin{align}
		R\rightarrow R-\beta,\qquad \beta \simeq 0.946,
	\end{align}
	where the value of $ \beta $ is determined by numerical fitting of $ \epsilon_0 $ (Fig.~\ref{fig:kelvin-epsilonzero}). The physical meaning of this replacement is as follows. Since the order parameter is suppressed near the boundary, the effective radius of the system becomes about a healing length shorter than that of the Neumann BC. See also Fig.~\ref{fig:kelzerodirichlet}. \\
	\indent The formula obtained here explains the numerical results very well for small wavenumbers in finite-size systems. However, we cannot take the limit $ R\to\infty $ in this expression. In the next subsection, we derive an interpolating formula valid even for $ R=\infty $. 
\subsection{Interpolating formula, derivation of $ \epsilon \sim -k^2\log k $}
	Now we consider the finite-wavenumber case of Eqs. (\ref{eq:albogou}) and (\ref{eq:albogov}). Since we are interested in the region such that $ kR \sim O(1) $, we expand  the wavenumber as $ k=\tilde{k}\alpha+\dotsb $. The energy and quasiparticle wavefunctions are expanded in the same way with the previous subsection: $ \epsilon=\epsilon_0+\alpha^2\epsilon_2+\dotsb,\ (u,v)=(u_0,v_0)+\alpha^2(u_2,v_2)+\dotsb$. Then, the zeroth-order equations are
	\begin{align}
		2(u_0+v_0)&=\epsilon_0u_0, \label{eq:kelintr01}\\
		-2(u_0+v_0)&=\epsilon_0v_0. \label{eq:kelintr02}
	\end{align}
	In order for these equations to have a nonvanishing solution,  $ \det\left( \begin{smallmatrix} 2+\epsilon_0 & 2 \\ 2 & 2-\epsilon_0 \end{smallmatrix} \right)=0 $ is necessary. Thus we obtain $ \epsilon_0=0 $ and $ u_0+v_0=0 $, which again gives the justification for the procedure (B) in Sec.~\ref{subsec:sketch}. 
	The second order equations are given by
	\begin{align}
		-u_0''-\frac{u_0'}{\xi}-\frac{u_0}{\xi^2}+\tilde{k}^2u_0+2(u_2+v_2)&=\epsilon_2u_0, \label{eq:kelintr03}\\
		v_0''+\frac{v_0'}{\xi}-\frac{3v_0}{\xi^2}-\tilde{k}^2v_0-2(u_2+v_2)&=\epsilon_2v_0. \label{eq:kelintr04}
	\end{align}
	Taking the sum of these two equations and using $ u_0+v_0=0 $, we obtain
	\begin{align}
		-u_0''-\frac{u_0'}{\xi}+\frac{u_0}{\xi^2}+\tilde{k}^2u_0=0,
	\end{align}
	which is just the modified Bessel differential equation. Thus the solution is given by $ u_0=c_1I_1(\tilde{k}\xi)+c_2K_1(\tilde{k}\xi) $. Again, following the procedure (B), imposing the Neumann BC $ \lim_{\xi\rightarrow 1}u'(\xi)=0 $, we obtain
	\begin{align}
		u_0&=\tilde{k}\left[ K_1(\tilde{k}\xi)+\chi(\tilde{k})I_1(\tilde{k}\xi) \right],\\
		\chi(\tilde{k})&:=\frac{K_0(\tilde{k})+K_2(\tilde{k})}{I_0(\tilde{k})+I_2(\tilde{k})}.
	\end{align}
	If we go back to the original variables $ r $ and $ R $, this solution can be rewritten as
	\begin{align}
		u_0=F_k(r):=k\left[ K_1(kr)+\chi(kR)I_1(kr) \right].
	\end{align}
	This $ F_k(r) $ has a few notable properties. If  $ R\ne\infty $, it has a Taylor series around $ k=0 $:
	\begin{align}
		F_k(r)&=\frac{1}{r}+\frac{r}{R^2}+\frac{k^2 r}{8}\left( -7+\frac{r^2}{R^2}+4\log\frac{r}{R} \right)+O(k^4),
	\end{align}
	which implies that the naive perturbation works out well if the system size is finite. 
	On the other hand, if $ R=\infty, $ the function $ \chi(\tilde{k}) $ has the asymptotic behavior 
	\begin{align}
		\chi(\tilde{k})=\begin{cases} \frac{2}{\tilde{k}^2}-\gamma-\frac{5}{4}-\log\frac{\tilde{k}}{2}+O(\tilde{k}^2) & (\tilde{k}\ll 1) \\ \pi \mathrm{e}^{-2\tilde{k}}[1+\tfrac{7}{4\tilde{k}}+O(\tilde{k}^{-2})] & (\tilde{k} \gg 1). \end{cases}
	\end{align}
	Hence, $ \lim_{R\rightarrow\infty}F_k(r) =k K_1(kr) $, which does not have a Taylor series since $ K_1(kr) $ includes the logarithmic term. Thus, we cannot use the naive perturbation theory in the infinite-size system. We mention that the solution $ u=K_1(kr) $ was also found by Pitaevskii \cite{Pitaevskii1961}.\\
	\indent Now, following the same procedure with the previous subsection, we modify this solution in order to avoid the artificial divergence at $ r=0 $. Namely, we use the following modified quasiparticle wavefunction:
	\begin{align}
		w_k:=\begin{pmatrix}u_k(r) \\ v_k(r)\end{pmatrix}=\frac{1}{\sqrt{2}}\begin{pmatrix} F_k(r)-\frac{1}{r}+\left( \frac{f_\infty(r)}{r}+f_\infty'(r) \right) \\ -F_k(r)+\frac{1}{r}-\left( \frac{f_\infty(r)}{r}-f_\infty'(r) \right)\end{pmatrix}. \label{eq:modifiedFksupp}
	\end{align}
	This expression just gives Eq.~(\ref{eq:kelFr}) up to a factor. If we set $ k=0 $ in this expression, we again obtain Eq.~(\ref{eq:kelmodzeromode}). \\
	\indent Let us calculate the eigenenergy  $ \epsilon_k $ of $ w_k $ by solving the Bogoliubov equation [Eq.~(\ref{eq:kelfundH00})]
	\begin{align}
		(H_0+\sigma k^2)w_k=\epsilon_kw_k.
	\end{align}
	Taking the $ \sigma $-inner product between this equation and $ w_0 $, we obtain
	\begin{align}
		\epsilon_k&=\epsilon_0+k^2\frac{(w_0,\sigma w_k)_\sigma}{(w_0,w_k)_\sigma} \nonumber \\
		&=\epsilon_0+k^2\frac{\int_0^Rr\mathrm{d}r\left[u_0(r)u_k(r)+v_0(r)v_k(r)\right]}{\int_0^Rr\mathrm{d}r\left[u_0(r)u_k(r)-v_0(r)v_k(r)\right]}.
	\end{align} 
	We already know $ \epsilon_0=-\frac{2}{R^2} $ [Eq.~(\ref{eq:kelepsilonzero})]. Let us calculate the inner products. We write
	\begin{align}
		\int_0^Rr\mathrm{d}r\left[u_0(r)u_k(r)-v_0(r)v_k(r)\right]&=I_1+I_2, \\
		\int_0^Rr\mathrm{d}r\left[u_0(r)u_k(r)+v_0(r)v_k(r)\right]&=I_3+I_4+I_5,
	\end{align}
	where
	\begin{align}
		I_1&=\int_0^R\mathrm{d}r f_\infty'(r) \left[ 2f_\infty(r)+\tfrac{2r^2}{R^2} \right],\\
		I_2&=\int_0^Rr\mathrm{d}r f_\infty'(r)[F_k(r)-F_0(r)],\\
		I_3&=\int_0^Rr\mathrm{d}r\left[u_0(r)^2+v_0(r)^2\right],\\
		I_4&=\int_0^Rr\mathrm{d}rF_0(r)[F_k(r)-F_0(r)],\\
		I_5&=\int_0^R\mathrm{d}r[f_\infty(r)-1][F_k(r)-F_0(r)].
	\end{align}
	The integrals $ I_1 $ and $ I_3 $ are $ k $-independent and already evaluated in the previous subsection [Eqs. (\ref{eq:w0sigmaw0}) and (\ref{eq:w0sigmsigmaaw0})]:
	\begin{align}
		I_1&=1+O\left( \frac{\log R}{R^2} \right),\\
		I_3&=\log R+\frac{5}{4}+\eta+O\left( \frac{\log R}{R^2} \right).
	\end{align}
	If we perform the order evaluation by regarding $ k=O(R^{-1}) $,  $ I_2 $ and $ I_5 $ are shown to be ignorable:
	\begin{align}
		I_2=O\left( \frac{(\log R)^2}{R^2} \right),\quad I_5 = O\left( \frac{(\log R)^2}{R^2} \right).
	\end{align}
	$ I_4 $ can be symbolically integrated as
	\begin{align}
		I_4&=\Bigg[ \chi(kR)\left( I_0(kr)+\frac{r^2I_2(kr)}{R^2} \right)\nonumber \\
		&\quad-\left(\log r+K_0(kr)\right)-\frac{r^2 K_2(kr)}{R^2}-\frac{r^2}{R^2}-\frac{r^4}{4R^4} \Bigg]_0^R \nonumber \\
		&=\frac{2}{k^2R^2}-\log\frac{kR}{2}-\frac{5}{4}-\gamma-\chi(kR),
	\end{align}
	where the behaviors $ K_0(kr)+\log r=-\gamma-\frac{\log k}{2}+O(r^2) $ and $ K_2(kr)=\frac{2}{k^2r^2}+O(1) $ are used.\\
	\indent Summarizing, the dispersion relation of the Kelvin mode is given by
	\begin{align}
		\epsilon_k=k^2\left( -\log\frac{k}{2}+\eta-\gamma-\chi(kR) \right). \label{eq:Kelvincasessuppmat00}
	\end{align}
	This formula includes the following two limiting cases:
	\begin{subnumcases}
		{\epsilon_k \simeq}
		\!\!\!\displaystyle-\frac{2}{R^2}+k^2\left(\log R+\frac{5}{4}+\eta\right) \!\!& $(kR\ll 1)$ \\ \!\!\displaystyle k^2\left(-\log\frac{k}{2}+\eta-\gamma\right) \!\!& $(R\rightarrow\infty)$. \label{eq:Kelvincasessuppmat}
	\end{subnumcases}
	The case $ kR \ll 1 $ revisits Eq.~(\ref{eq:kelzeronaiv301}) with (\ref{eq:kelzeronaiv302}). The latter case gives the non-integer dispersion $ \epsilon\sim-k^2\log k $, which was first shown in Ref.~\onlinecite{Pitaevskii1961}. 
	Taking the limit $ R\rightarrow\infty $ in Eq.~(\ref{eq:modifiedFksupp}), the quasiparticle wavefunction of Kelvin modes in the infinite system becomes
	\begin{align}
		\begin{pmatrix}u_k(r) \\ v_k(r)\end{pmatrix}=\frac{1}{\sqrt{2}}\begin{pmatrix} kK_1(kr)-\frac{1}{r}+\frac{f_\infty(r)}{r}+f_\infty'(r) \\ -kK_1(kr)+\frac{1}{r}-\frac{f_\infty(r)}{r}+f_\infty'(r)\end{pmatrix}. 
	\end{align}
	\indent Finally, we would like to give a few perspectives on the higher-order corrections of the dispersion relation Eq.~(\ref{eq:Kelvincasessuppmat00}) and its infinite limit Eq.~(\ref{eq:Kelvincasessuppmat}). In deriving this formula, we have ignored the terms $ I_2 $ and $ I_5 $ by assuming $ k=O(R^{-1}) $. Since they vanish at $ k=0 $, this ignoring is not bad even in the infinite system $ R=\infty $, if $ k $ is small. Indeed, the numerical result with $ R=100 $ given in Fig.~\ref{fig:kelvin-dispersion} shows that the formula (\ref{eq:Kelvincasessuppmat}) is good for $ 0\le k\lesssim 0.1 $. However, if we are interested in the next leading order term of the formula (\ref{eq:Kelvincasessuppmat00}), we must include contributions from $ I_2 $ and $ I_5 $. The emergence of $ O\big((\log R)^2/R^2\big) $ terms implies that, if these terms are treated with mathematical care, they will become of order $ O\big(k^2(\log k)^2\big) $, meaning that the next leading term of the dispersion relation Eq.~(\ref{eq:Kelvincasessuppmat}) would be given by $ k^4(\log k)^2 $. However, at this time, we do not have a derivation for this conjecture and a possible finite-size generalization. This is left to be an open problem.
\section{Detailed derivation --- ripplons}\label{sec:ripplon}
	In this section we provide the detailed derivations of analytical formulas for ripplons presented in Sec.~\ref{sec:mainresult}.
\subsection{Fundamental equations and ground states in 1D systems}\label{subsec:ripGS}
	We first consider the ground state of the one-dimensional system with length $ 2L $ 
	\begin{align}
		H&=\int_{-L}^L\mathrm{d}x \Bigg(\frac{|\partial_x\psi_1|^2}{2m_1}+\frac{|\partial_x\psi_2|^2}{2m_2}\nonumber \\
		&\quad\qquad+g_{11}|\psi_1|^4+2g_{12}|\psi_1|^2|\psi_2|^2+g_{22}|\psi_2|^4\Bigg)
	\end{align}
	with fixed particle numbers $ N_i=\int\mathrm{d}x|\psi_i|^2 $,\ $ i=1,2 $. Though the result of this problem is well-known\cite{PhysRevA.58.4836,PethickSmith}, we review it in order to introduce the variable $ d $ [Eq.~(\ref{eg:defposiDW})], having the meaning of the position of the DW. The discussion given below holds regardless of whether the BC at $ x=\pm L $ is of Dirichlet or Neumann. \\
	\indent Let us assume that the system length $ 2L $ is sufficiently large compared to the typical healing length of the order parameters and hence the energies of bulk condensates are much larger than those of surfaces and boundaries. (We can introduce four kinds of healing lengths in this system 
as seen in Appendix~\ref{app:miscripplon}.) Assume that two condensates $ \psi_1 $ and $ \psi_2 $ are separated, and $ \psi_{1(2)} $ occupies the left (right) side of the box with length $ L_{1(2)} $, where $ L_1+L_2=2L $. Then, the energy of this state is given by
	\begin{align}
		H_{\text{separated}}=g_{11}\frac{N_1^2}{L_1}+g_{22}\frac{N_2^2}{2L-L_1}.
	\end{align}
	Minimization of $ H_{\text{separated}} $ with respect to $ L_1 $ yields
	\begin{align}
		\frac{L_i}{2L}&=\frac{\sqrt{g_{ii}}N_i}{\sqrt{g_{11}}N_1+\sqrt{g_{22}}N_2},\quad i=1,2,\\
		H_{\text{separated}}&=\frac{(\sqrt{g_{11}}N_1+\sqrt{g_{22}}N_2)^2}{2L}.
	\end{align}
	On the other hand, as another ansatz, the energy of the uniform mixture of $ \psi_1 $ and $ \psi_2 $ is given by
	\begin{align}
		H_{\text{mixed}}=\frac{g_{11}N_1^2+g_{22}N_2^2+2g_{12}N_1N_2}{2L},
	\end{align}
	which does not have an additional parameter to be optimized. The energy difference between these two states is given by
	\begin{align}
		H_{\text{mixed}}-H_{\text{separated}}=\frac{N_1N_2(g_{12}-\sqrt{g_{11}g_{22}})}{L}.
	\end{align}
	Thus, if $ g_{12}>\sqrt{g_{11}g_{22}} $, the ground state is given by the state such that $ \psi_1 $ and $ \psi_2 $ are separated.\\ 
	\indent Henceforth we only consider the separated case. The densities of these condensates are given by
	\begin{align}
		\rho_i=\frac{N_i}{L_i}=\frac{\sqrt{g_{11}}N_1+\sqrt{g_{22}}N_2}{2L\sqrt{g_{ii}}},\quad i=1,2.
	\end{align}
	If we introduce
	\begin{align}
		p:=\frac{\sqrt{g_{11}}N_1+\sqrt{g_{22}}N_2}{2L},
	\end{align}
	the densities can be rewritten as
	\begin{align}
		p=\sqrt{g_{11}}\rho_1=\sqrt{g_{22}}\rho_2. \label{eq:ripplonp2}
	\end{align}
	This relation also holds in the infinite-size system due to the momentum conservation law (see Appendix~\ref{app:miscripplon}).  The position of the DW is given by
	\begin{align}
		d:=L_1-L=\frac{L(\sqrt{g_{11}}N_1-\sqrt{g_{22}}N_2)}{\sqrt{g_{11}}N_1+\sqrt{g_{22}}N_2}. \label{eg:defposiDW}
	\end{align}
	We can use  $ p $ and $ d $ as system parameters instead of $ N_1 $ and $ N_2 $. The relation between them are
	\begin{align}
		N_1=\frac{p(L+d)}{\sqrt{g_{11}}},\ N_2=\frac{p(L-d)}{\sqrt{g_{22}}}. \label{eq:N1N2expression}
	\end{align}
	Henceforth we regard $ \psi_i $'s as functions of these parameters instead of $ N_1 $ and $ N_2 $, that is, they are considered as a function $ \psi_i=\psi_i(x,p,d) $. 
	If the system length $ 2L $ is sufficiently large and the DW is located far from the boundary (i.e., $ |d\pm L| $ is much larger than the typical healing length), changing $ d $ with fixed $ p $ implies a smooth sliding of the DW almost without changing the profiles of $ \psi_1,\psi_2 $ far from the DW. 
	If $ g_{11}=g_{22} $, the story becomes a little  simpler; since $ p\propto N_1+N_2 $ and $ d\propto N_1-N_2 $, the sliding of the DW occurs by changing the imbalance of the particle numbers $ N_1-N_2 $ with fixing the total number $ N_1+N_2 $. In the general case $ g_{11}\ne g_{22} $, however, fixing $ p $ does not mean fixing the total particle number. \\ 
	\indent From the above physical interpretation, the differentiation with respect to $ d $ with fixed  $ p $ is approximately given by
	\begin{align}
		\frac{\partial }{\partial d} \simeq \begin{cases} -\frac{\partial }{\partial x} & (x\simeq d) \\ 0 & (|x-d|\gg\xi), \end{cases} \label{eq:dderivative}
	\end{align}
	with $ \xi $ being the typical healing length. In particular, if we take the infinite-size limit, we have
	\begin{align}
		\lim_{L\rightarrow\infty}\frac{\partial }{\partial d}=-\frac{\partial }{\partial x}. \label{eq:ripddertoxder}
	\end{align}
\subsection{SSB-originated zero mode solutions and overview of calculation}
	Now let us consider a three-dimensional system. We consider the system such that the length with respect to the $ x $-direction is $ 2L $ and those with respect to the $ y $- and $ z $-directions are infinite: $ L_y=L_z=\infty $. As shown below, if $ L_y=L_z=\infty $, the Bogoliubov equation has the complex eigenvalue. In other words, the system has unstable modes. However, the wavenumbers of unstable modes are shown to be exponentially small $ k \sim \mathrm{e}^{-\alpha L} $, therefore we can easily eliminate these unstable modes through discretization of wavenumbers, which is realized by modifying $ L_y,L_z $ to be very large but finite sizes. \\
	\indent Let us consider the GP and Bogoliubov equations. Assuming  translationally-invariant configurations along the $ y $- and $ z $- directions, the GP equation is reduced to
	\begin{align}
		\left(-\mu_1-\tfrac{1}{2m_1}\partial_x^2+2g_{11}|\psi_1|^2+2g_{12}|\psi_2|^2\right)\psi_1&=0, \\
		\left(-\mu_2-\tfrac{1}{2m_2}\partial_x^2+2g_{21}|\psi_1|^2+2g_{22}|\psi_2|^2\right)\psi_2&=0.
	\end{align}
	For the DW solution,  $ \psi_1 $ and $ \psi_2 $ can be taken as real-valued functions up to overall phase factors. 
	The chemical potentials are the functions of system parameters: $ \mu_i=\mu_i(p,d) $, and they are determined via the condition $ \int\mathrm{d}x|\psi_i|^2=N_i $. If $ L $ is large, they are almost the same with those of the infinite-size system: $ \mu_i \simeq 2g_{ii}\rho_i=2\sqrt{g_{ii}}p $ (see Appendix~\ref{app:miscripplon}). Therefore, the $ d $-dependence of $ \mu_i $'s is expected to be very small for large $ L $. \\
	\indent For the Bogoliubov equation, assuming the plane-wave solution in the $ y $- and $ z $- directions, we set $ u_i(x,y,z)=u_i(x)\mathrm{e}^{\mathrm{i}(k_yy+k_zz)},\ v_i(x,y,z)=v_i(x)\mathrm{e}^{\mathrm{i}(k_yy+k_zz)},\ i=1,2 $. We then obtain
	\begin{align}
		(H_0+M_0 k^2)\begin{pmatrix}u_1 \\ u_2 \\ v_1 \\ v_2 \end{pmatrix}=\epsilon \begin{pmatrix}u_1 \\ u_2 \\ v_1 \\ v_2 \end{pmatrix}, \label{eq:Bogoripp}
	\end{align}
	where $ k=\sqrt{k_y^2+k_z^2},\ M_0=\operatorname{diag}(\frac{1}{2m_1},\frac{1}{2m_2},\frac{-1}{2m_1},\frac{-1}{2m_2}) $, and 
	\begin{align}
		H_0=\begin{pmatrix} F_0 & G_0 \\ -G_0^* & -F_0^* \end{pmatrix}
	\end{align}
	with
	\begin{align}
		F_0&=\operatorname{diag}\left(-\tfrac{\partial_x^2}{2m_1}-\mu_1,-\tfrac{\partial_x^2}{2m_2}-\mu_2\right)\nonumber \\
		&\quad+\begin{pmatrix} 4g_{11}|\psi_1|^2+2g_{12}|\psi_2|^2 & 2g_{12}\psi_1\psi_2^* \\ 2g_{12}\psi_1^*\psi_2 & 4g_{22}|\psi_2|^2+2g_{12}|\psi_1|^2 \end{pmatrix}, \\
		G_0&=\begin{pmatrix}2g_{11}\psi_1^2 & 2g_{12}\psi_1\psi_2 \\ 2g_{12}\psi_1\psi_2 & 2g_{22}\psi_2^2\end{pmatrix}.
	\end{align}
	Note that the kinetic energy term is not $ \sigma=\operatorname{diag}(1,1,-1,-1) $, because the masses are generally different: $ m_1\ne m_2 $. The $ \sigma $-inner product between two quasiparticle wavefunctions $ w_i=(u_{i1},u_{i2},v_{i1},v_{i2})^T,\ i=1,2 $ is defined by 
	\begin{align}
		&(w_1,w_2)_\sigma=\int \mathrm{d}x\left( u_{11}^*u_{21}+u_{12}^*u_{22}-v_{11}^*v_{21}-v_{12}^*v_{22} \right). \label{eq:rippsigmaprd}
	\end{align} 
	$ H_0 $ and $ M_0 $ satisfy the ``Bogoliubov-hermitian'' property \cite{DTMN}: 
	\begin{align}
		(x,H_0y)_\sigma&=(H_0x,y)_\sigma, \label{eq:rippH0BH} \\
		(x,M_0y)_\sigma&=(M_0x,y)_\sigma.
	\end{align}
	\indent Let us discuss SSB-originated zero-mode solutions \cite{DTMN}. In the infinite-size system, if $ (\psi_1(x,y,z),\psi_2(x,y,z)) $ is a solution of the GP equation,  $ (\mathrm{e}^{\mathrm{i}\theta_1}\psi_1(x+x_0,y,z),\mathrm{e}^{\mathrm{i}\theta_2}\psi_2(x+x_0,y,z)) $ is also a solution. By differentiating the GP equation with respect to $ \theta_1,\,\theta_2 $ and $ x_0 $, we have the following zero-mode solutions: 
	\begin{align}
		w_1=\begin{pmatrix}\psi_1 \\ 0 \\ -\psi_1^* \\ 0 \end{pmatrix},\quad w_2=\begin{pmatrix}0 \\ \psi_2 \\ 0 \\ -\psi_2^* \end{pmatrix},\quad w_{\text{trans}}=\frac{\partial }{\partial x}\begin{pmatrix}\psi_1 \\ \psi_2 \\ \psi_1^* \\ \psi_2^*\end{pmatrix}.
	\end{align}
	However, if we consider a finite-size system, only $ w_1 $ and $ w_2 $ are exact zero-mode solutions and $ w_{\text{trans}} $ is no longer a solution since the translational symmetry is absent. 
	In the finite-size system, the generalized eigenvector $ z_d $, derived in the next subsection, plays an alternative role to $ w_{\text{trans}} $. \\ 
	\indent Since these two modes are  $ \sigma $-orthogonal to each other $(w_1,w_2)_\sigma=0$,  we conclude that the system has two type-I NGMs and no type-II NGM appears by
following the general theory constructed in Ref.~\onlinecite{DTMN}. 
At first glance, this fact would seem contradictory to the fact 
that the ripplon has a type-II dispersion in a finite-size system \cite{PhysRevA.88.043612,DTMN}. This apparent paradox can be resolved in the following way: the gapless mode corresponding to the ripplon indeed has a linear dispersion $ \epsilon = a k $ in finite-size systems. However, the coefficient $ a $ is an exponentially small \textit{complex} number. If we ignore this exponentially small region $ k\lesssim O(\mathrm{e}^{-L/\xi}) $, the dispersion relation for $ k\lesssim O(L^{-1}) $ is well described by $ \epsilon \sim \sqrt{L}k^2 $, as shown in Refs.~\onlinecite{PhysRevA.88.043612,DTMN}. Furthermore, if $ k $ becomes a little larger, the dispersion relation becomes $ \epsilon \sim k^{3/2} $. 
	These three different behaviors in different wavenumber scales will be solely explained by one formula in Eqs.~(\ref{eq:ripdispmod}) and (\ref{eq:ripak}), which are the goal of this section.  
	Henceforth, we solve the Bogoliubov equation in the three ways shown in Table~\ref{ta:rippaprox} to derive the above-mentioned three behaviors. Even though the last method provided in Subsec.~\ref{subsec:rippinterpol} gives the most general and important result, the former methods treated in Subsecs.~\ref{subsec:ripnaive} and \ref{subsec:ripptwostate} are necessary to formulate the last method. So we need all three formulations.
	\begin{table}[tb]
		\begin{center}
		\caption{\label{ta:rippaprox} A list of approximations and derivable dispersion relations for the ripplons in finite-size systems. The naive perturbation, two-state approximation, and $ k $-dependent two-state approximation are discussed in Subsecs.~\ref{subsec:ripnaive}, \ref{subsec:ripptwostate}, and \ref{subsec:rippinterpol}, respectively.}
		{\small
		\begin{tabular}{r|c|c|c|}
			& $\epsilon\propto\mathrm{i}ck$ & $\epsilon\propto\!\sqrt{L}k^2$ & $\epsilon\propto k^{3/2}$ \\
		\hline
		Naive perturbation & \checkmark & & \\
		\hline
		Two-state approximation & \checkmark & \checkmark & \\
		\hline
		$ k $-dependent two-state approximation & \checkmark & \checkmark & \checkmark \\
		\hline
		\end{tabular}
		}
		\end{center}
	\end{table}
\subsection{Naive perturbation --- type-I complex dispersion}\label{subsec:ripnaive}
	\indent We first solve the Bogoliubov equation by a naive perturbation theory and find the complex-coefficient type-I dispersion. \\
	\indent Since $ w_1 $ and $ w_2 $ are the seeds of type-I NGMs, there must exist generalized eigenvectors satisfying $ H_0z_i\propto w_i $ according to Ref.~\onlinecite{DTMN}. Such vectors can be found by differentiating the GP equation with respect to the system parameters \cite{DTMN,TakahashiPhysD}. The differentiation with respect to $ p $ and $ d $ yields
	\begin{align}
		H_0z_p&=\mu_{1p}w_1+\mu_{2p}w_2, \\
		H_0z_d&=\mu_{1d}w_1+\mu_{2d}w_2, \label{eq:ripzd}
	\end{align}
	where
	\begin{align}
		z:=\begin{pmatrix} \psi_1 \\ \psi_2 \\ \psi_1^* \\ \psi_2^* \end{pmatrix},\quad  z_p:=\frac{\partial z}{\partial p},\quad z_d:=\frac{\partial z}{\partial d}, \\
		\mu_{ip}:=\frac{\partial \mu_i}{\partial p},\quad \mu_{id}:=\frac{\partial \mu_i}{\partial d},\quad i=1,2.
	\end{align}
	Let us define the following notation for later convenience
	\begin{align}
		[A,B]_{pd}:=\frac{\partial A}{\partial p}\frac{\partial B}{\partial d}-\frac{\partial B}{\partial p}\frac{\partial A}{\partial d}.
	\end{align}
	Then, if we introduce
	\begin{align}
		z_1=\frac{[z,\mu_2]_{pd}}{[\mu_1,\mu_2]_{pd}},\quad z_2=\frac{[\mu_1,z]_{pd}}{[\mu_1,\mu_2]_{pd}},
	\end{align}
	they satisfy
	\begin{align}
		H_0z_i=w_i,\quad i=1,2. \label{eq:zimu3}
	\end{align}
	As already mentioned, if the system length $ L $ is sufficiently large, $ \mu_i $ can be approximated by those of infinite-size systems: $ \mu_i\simeq 2g_{ii}\rho_i\simeq2\sqrt{g_{ii}}p $. (See Appendix~\ref{app:miscripplon}.) Therefore,
	\begin{align}
		\mu_{ip}\simeq 2\sqrt{g_{ii}},\quad \mu_{id}\simeq 0,\quad i=1,2. \label{eq:muipmuid}
	\end{align}
	This implies that $ \mu_{id} $ vanishes if we only take the leading order. 
A rigorous evaluation of $ \mu_{1d}, \mu_{2d} $ is not easy, but the typical behavior is given by 
	\begin{align}
		\mu_{id} \sim L\mathrm{e}^{-\alpha L/\xi},\ \mu_{1d}<0,\ \mu_{2d}>0, \label{eq:muidbehav}
	\end{align}
	where $ \xi $ is the typical healing length of order parameters and  $ \alpha $ is an $ O(1) $ constant.  For the special case $ g_{12}=+\infty $, we can rigorously derive the behavior in Eq.~(\ref{eq:muidbehav}),  because the two condensates are completely separated and hence the GP equation reduces to that of a single-component BEC. See Appendix~\ref{app:evaluatemuderiv}. We can also find similar behaviors for finite $ g_{12} $ from numerics. As we see below, these small $ \mu_{id} $'s cause a very narrow complex eigenvalue region in the dispersion relation. Since $ \mu_{id} $'s are very small, we often ignore higher-order terms of $ \mu_{id} $'s in the following calculation. \\
	\indent Because of Eqs.~(\ref{eq:ripzd}) and (\ref{eq:muidbehav}), the generalized eigenvector $ z_d $ is an ``almost'' zero-mode solution if $ L $ is large.  
	In particular, using Eq.~(\ref{eq:ripddertoxder}), it exactly reduces to the zero mode solution due to the translational symmetry breaking in the infinite-size limit: 
	\begin{align}
		z_d \rightarrow -w_{\text{trans}}=-\frac{\partial }{\partial x}\begin{pmatrix} \psi_1 \\ \psi_2 \\ \psi_1^* \\ \psi_2^* \end{pmatrix} \quad (L\rightarrow\infty). \label{eq:zdtowtrans}
	\end{align}
	This relation implies that $ z_d $ plays an alternative role to $ w_{\text{trans}} $ in finite-size systems. \\ 
	\indent Let us derive the eigenvectors and eigenvalues of the Bogoliubov equations (\ref{eq:Bogoripp}) by solving it perturbatively \cite{DTMN}.  Let us look for the eigenvector and eigenvalue by the expansion
	\begin{align}
		\zeta&=\zeta_0+\zeta_1k+\zeta_2k^2+\dotsb,\\
		\epsilon&=\epsilon_1 k+\epsilon_2k^2+\dotsb
	\end{align}
	with
	\begin{align}
		\zeta_0&=a_1w_1+a_2w_2, \\
		\zeta_1&=b_1z_1+b_2z_2.
	\end{align}
	The zeroth order equation $ H_0\zeta_0=0 $ holds identically. From the first-order equation $ H_0\zeta_1=\epsilon_1\zeta_0 $, we obtain 
	\begin{align}
		b_1=\epsilon_1a_1,\quad b_2=\epsilon_1a_2.
	\end{align}
	The second-order equation is given by $ M_0\zeta_0+H_0\zeta_2=\epsilon_2\zeta_0+\epsilon_1\zeta_1 $. Taking the $ \sigma $-inner product between this equation and $ w_i $ gives
	\begin{align}
	W\begin{pmatrix}a_1 \\ a_2 \end{pmatrix}=\epsilon_1^2G\begin{pmatrix}a_1 \\ a_2 \end{pmatrix}, \label{eq:rippniv}
	\end{align}
	where $ W $ and $ G $ are  $ 2\times 2 $ matrices whose components are defined by
	\begin{align}
		[W]_{ij}=(w_i,M_0w_j)_\sigma,\quad [G]_{ij}=(w_i,z_j)_\sigma.
	\end{align}
	They can be calculated as 
	\begin{align}
		W&=\begin{pmatrix} N_1/m_1 & 0 \\ 0 & N_2/m_2 \end{pmatrix}, \label{eq:rippW2} \\
		G&=\frac{1}{[\mu_1,\mu_2]_{pd}}\begin{pmatrix} [N_1,\mu_2]_{pd} & [\mu_1,N_1]_{pd} \\ [N_2,\mu_2]_{pd} & [\mu_1,N_2]_{pd} \end{pmatrix}. \label{eq:rippgramlike}
	\end{align}
	Note that the entries of $ G $ can be also written as $ [G]_{ij}=(H_0z_i,z_j)_\sigma $ by Eq.~(\ref{eq:zimu3}), and hence $ G $ is hermitian due to Eq.~(\ref{eq:rippH0BH}). Furthermore, Eq.~(\ref{eq:rippgramlike}) shows that $ G $ is real, hence $ G $ is a real-symmetric matrix. We thus obtain the following relation between parameter derivatives:
	\begin{align}
		[\mu_1,N_1]_{pd}=[N_2,\mu_2]_{pd}. \label{eq:paraidentty}
	\end{align}
	Using (\ref{eq:N1N2expression}), it can be rewritten as 
	\begin{align}
		\mu_{1p}\sqrt{g_{22}}-\mu_{2p}\sqrt{g_{11}}=\frac{\sqrt{g_{22}}(L+d)\mu_{1d}+\sqrt{g_{11}}(L-d)\mu_{2d}}{p}, \label{eq:paraderid}
	\end{align}
	which shows that the parameter derivatives $ \mu_{1p},\mu_{1d},\mu_{2p}, $ and $ \mu_{2d} $ are, in fact, not independent. The identity between parameter derivatives similar to Eq. (\ref{eq:paraidentty}) was also reported in Appendix A of Ref.~\onlinecite{TakahashiPhysD}. \\ 
	\indent By solving the eigenvalue problem (\ref{eq:rippniv}) up to leading order for $ \mu_{1d} $ and $ \mu_{2d} $, we obtain the following result: \\
	\indent The dispersion relation and eigenvector corresponding to the Bogoliubov phonon are given by
	\begin{align}
		\epsilon&=c_{\text{ph}}k+O(k^2),\\
		\zeta&=w_{\text{ph}}+z_{\text{ph}}c_{\text{ph}}k+O(k^2) \\
	\intertext{with}
		c_{\text{ph}}^2&=\frac{g_{11}\rho_1}{m_1}\left( 1+\frac{d}{L} \right)+\frac{g_{22}\rho_2}{m_2}\left( 1-\frac{d}{L} \right), \label{eq:ripnivph}\\
		w_{\text{ph}}&=\sqrt{g_{11}}w_1+\sqrt{g_{22}}w_2,\ z_{\text{ph}}=\tfrac{1}{2}z_p.
	\end{align}
	Here, ``ph'' means the phonon. Strictly speaking, the first order eigenvector $ z_{\text{ph}} $ may include $ z_d $, but we ignore it because it is not important in order for the first-order equation $ H_0\zeta_1=\epsilon_1\zeta_0 $ to be satisfied up to $ O(\mu_{id}) $. \\
	\indent The dispersion relation and eigenvector corresponding to ripplons are given by
	\begin{align}
		\epsilon&=c_{\text{rip}}k+O(k^2),\\
		\zeta&=w_{\text{rip}}+z_{\text{rip}}c_{\text{rip}}k+O(k^2) \\
		\intertext{with}
		c_{\text{rip}}^2&=\frac{(L^2-d^2)(\rho_1\mu_{1d}-\rho_2\mu_{2d})}{m_1\rho_1(L-d)+m_2\rho_2(L+d)}+O(\mu_{id}^2),\label{eq:ripnivrip}\\
		w_{\text{rip}}&=\left( 1-\frac{d}{L} \right)m_1w_1-\left( 1+\frac{d}{L} \right)m_2w_2, \label{eq:wrippzd}\\
		z_{\text{rip}}&=\frac{L^2-d^2}{Lc_{\text{rip}}^2}z_d+O(\mu_{id}^0). \label{eq:zrippzd}
	\end{align}
	Since $ \rho_1\mu_{1d}-\rho_2\mu_{2d} $ is exponentially small and negative [Eq.~(\ref{eq:muidbehav})], $ c_{\text{rip}} $ is pure imaginary. Therefore, this dispersion relation represents the existence of unstable modes in a very narrow wavenumber region.  
\subsection{Two-state approximation --- quadratic dispersion}\label{subsec:ripptwostate}
	In the above naive perturbation method, we cannot obtain the dispersion relations of ripplons in the finite-size system  $ \epsilon \sim \sqrt{L}k^2 $. 
	In this subsection, we give a little better treatment to derive this. If the eigenenergy of the Bogoliubov equation is sufficiently small, only ripplon excitations exist. So, the eigenvector is well approximated by a linear combination of two vectors, $ w_{\text{rip}} $ and $ z_{\text{rip}} $. Using this fact, we solve the Bogoliubov equation non-perturbatively under the approximation such that the state space is spanned only by these two vectors. The result contains not only the previous complex-coefficient linear dispersion but also the $ \sqrt{L}k^2 $ behavior. However, even in this treatment, we cannot obtain the dispersion relation and the eigenvector allowing to take the limit $ L\rightarrow\infty $. The final goal is given in the next subsection. \\
	\indent Let us solve the Bogoliubov equation
	\begin{align}
		(H_0+M_0k^2)\zeta=\epsilon\zeta \label{eq:rippbgtw}
	\end{align}
	with the assumption that the eigenstate is given by the linear combination of the above two vectors:
	\begin{align}
		\zeta=\alpha w_{\text{rip}}+\beta z_d. \label{eq:twostateansatz}
	\end{align}
	Here, we use $ z_d $ instead of $ z_{\text{rip}} $ as a basis vector, since $ z_{\text{rip}}\propto z_d $ up to leading order with respect to $ \mu_{id} $'s [Eq.~(\ref{eq:zrippzd})]. Different from the previous subsection, the coefficients $ \alpha $ and $ \beta $ are now $ k $-dependent. Taking the $ \sigma $-inner product between Eq.~(\ref{eq:rippbgtw}) and $ w_{\text{rip}} $, $ z_d $, we obtain the  $ 2\times 2 $ matrix equation
	\begin{align}
		\begin{pmatrix} -\epsilon & \frac{Lc_{\text{rip}}^2}{L^2-d^2}+k^2\frac{(z_d,M_0z_d)_\sigma}{(w_{\text{rip}},z_d)_\sigma} \\ k^2\frac{(w_{\text{rip}},M_0w_{\text{rip}})_\sigma}{(w_{\text{rip}},z_d)_\sigma} & -\epsilon \end{pmatrix}\begin{pmatrix}\alpha \\ \beta \end{pmatrix}=0, \label{eq:riptsace}
	\end{align}
	where we have used $ (z_i,M_0w_j)_\sigma=0 $, $ (w_i,M_0w_j)_\sigma=\delta_{ij}\frac{N_i}{m_i} $ for $ i,j=1,2 $ and $ H_0z_{\text{rip}}=w_{\text{rip}} \ \leftrightarrow \ H_0z_d=\frac{Lc_{\text{rip}}^2}{L^2-d^2}w_{\text{rip}} $. 
	Let us introduce the notation
	\begin{align}
		T_0:=\frac{(z_d,M_0z_d)_\sigma}{2}=\int_{-L}^L\mathrm{d}x\left( \frac{|\partial_d\psi_1|^2}{2m_1}+\frac{|\partial_d\psi_2|^2}{2m_2} \right),
	\end{align}
	which represents the kinetic energy of the DW.  By virtue of Eq.~(\ref{eq:dderivative}), the  $ d $-derivative takes up only the gradient energy of the DW, and it ignores the gradient energy near the boundaries $ x=\pm L $. This means that the leading value of $ (z_d,M_0z_d)_\sigma $ does not depend on a choice of the BC for sufficiently large $ L $, and hence it can be approximated by the kinetic energy of the DW in the infinite-size system:
	\begin{align}
		T_0 \simeq \int_{-\infty}^\infty\mathrm{d}x\left( \frac{|\partial_x\psi_1|^2}{2m_1}+\frac{|\partial_x\psi_2|^2}{2m_2} \right), \label{eq:T0inf}
	\end{align}
	where we should consider $ \psi_1 $ and $ \psi_2 $ of the infinite-size system when we use Eq.~(\ref{eq:T0inf}). 
	Then, solving Eq.~(\ref{eq:riptsace}) yields the dispersion relation and the eigenvector
	\begin{align}
		\epsilon^2&=A_0k^4+c_{\text{rip}}^2k^2=A_0k^2(k^2-k_c^2), \\ 
		\zeta&=\epsilon w_{\text{rip}}+\frac{L^2-d^2}{L}k^2z_d, \\
		A_0&:=\frac{2(L^2-d^2)T_0}{m_1\rho_1(L-d)+m_2\rho_2(L+d)}, \label{eq:ripa0} \\
		k_c&:=\sqrt{-c_{\text{rip}}^2/A_0}=\sqrt{\frac{\rho_2\mu_{2d}-\rho_1\mu_{1d}}{2T_0}}, \label{eq:ripkc}
	\end{align}
	respectively. 
Note that $ A_0 = O(L) $ and $ k_c $ is positive and of order $ O(\sqrt{L}\mathrm{e}^{-\alpha L/2\xi}) $ as a result of Eq.~(\ref{eq:muidbehav}). The very narrow region $ 0\le k \le k_c $ gives the unstable modes. If the physical parameters of $ \psi_1 $ and $ \psi_2 $ are symmetric, i.e.,  $ m_1=m_2,\ d=0,\ \rho_1=\rho_2,\ \mu_{2d}=-\mu_{1d}  $, it reduces to 
	\begin{align}
		\epsilon^2=\frac{LT_0}{m_1\rho_1}k^2(k^2-k_c^2), \quad k_c=\sqrt{\frac{-\rho_0\mu_{1d}}{T_0}}. \label{eq:ripkcsym}
	\end{align}
	If the narrow complex region is ignored, it gives $ \epsilon =\sqrt{\frac{LT_0}{m_1\rho_1}}k^2 $, as Ref.~\onlinecite{DTMN}. (Note that the mass is taken as $ 2m_1=1 $ in Ref.~\onlinecite{DTMN}.) 
\subsection{$ k $-dependent two-state approximation --- interpolating formula}\label{subsec:rippinterpol}
	The approximations used so far could not produce dispersion relations and eigenvectors which allow to take the limit $ L\rightarrow\infty $. To accomplish this, 
	let us construct a modified quasiparticle wavefunction including the asymptotic behavior far from the DW, corresponding to the general procedure (B) in Sec.~\ref{subsec:sketch}. Let us consider a uniform region $ -L+\xi\lesssim x\lesssim d-\xi $ so that the approximate expression $ \psi_1=\sqrt{\rho_1}\mathrm{e}^{\mathrm{i}\theta_1}=\text{const.} $ and $ \psi_2=0 $ can be well applied. Here $ \xi $ is a typical healing length of the condensates. We further introduce the notation $ F_1=u_1\mathrm{e}^{-\mathrm{i}\theta_1}-v_1\mathrm{e}^{\mathrm{i}\theta_1},\ G_1=u_1\mathrm{e}^{-\mathrm{i}\theta_1}+v_1\mathrm{e}^{\mathrm{i}\theta_1} $. Then, in this uniform region, the Bogoliubov equation can be written 
approximately as
	\begin{align}
		\frac{-\partial_x^2+k^2}{2m_1}F_1&=\epsilon G_1, \\
		\left( \frac{-\partial_x^2+k^2}{2m_1}+4g_{11}\rho_1 \right)G_1&=\epsilon F_1, \\
		\left( \frac{-\partial_x^2+k^2}{2m_2}+2(g_{12}\rho_1-g_{22}\rho_2) \right)u_2&=\epsilon u_2, \\
		-\left( \frac{-\partial_x^2+k^2}{2m_2}+2(g_{12}\rho_1-g_{22}\rho_2) \right)v_2&=\epsilon v_2.
	\end{align}
	Let us find a solution under the approximation such that we ignore functions whose decay rates are comparable with the healing lengths of condensates. (See Appendix~\ref{app:miscripplon} for expressions of the healing lengths.) We are interested in the wavenumber of order $ k\sim O(L^{-1}) $. Correspondingly we assume $ \epsilon \sim \sqrt{L}k^2 \sim O(L^{-3/2}) $. In this approximation, $ u_2 $ and $ v_2 $ are ignorable, because if we consider the solution $ u_2,v_2\propto \mathrm{e}^{\pm lx} $, we obtain $ l=\big(\kappa_{\text{DW2}}^2+k^2\pm2m_2\epsilon\big)^{1/2}= \kappa_{\text{DW2}}+O(L^{-3/2}) $, 
	where $ \kappa_{\text{DW2}} $ is defined in Eq.~(\ref{eq:apphldw2}). We thus set $ u_2=v_2=0 $. As for $ F_1 $ and $ G_1 $, if we assume $ (F_1,G_1)\propto \mathrm{e}^{\pm lx} $, we obtain  $ l=\kappa_1+O(L^{-1}) $ and $ l=k+O(L^{-2}) $, where $ \kappa_1 $ is defined in Eq.~(\ref{eq:apphl1}). The former solution is ignorable. The latter solution can contribute and the corresponding approximate eigenvector is given by
	\begin{align}
		\begin{pmatrix}F_1 \\ G_1 \end{pmatrix}=\begin{pmatrix} 1+O(L^{-3}) \\ O(L^{-3/2}) \end{pmatrix}\mathrm{e}^{\pm kx}
	\end{align}
	Thus, we can set $ F_1=\mathrm{e}^{\pm kx} $ and $ G_1=0 $, implying that the density fluctuation is ignorable, as stated in the procedure (B) of Sec.~\ref{subsec:sketch}.  Moreover, following the procedure (B), we impose the Neumann BC at $ x=-L $. Then, we have
	\begin{align}
		u_1\mathrm{e}^{-\mathrm{i}\theta_1}=-v_1\mathrm{e}^{\mathrm{i}\theta_1}=\cosh k(x+L),\quad u_2=v_2=0
	\end{align}
	for the region $ -L+\xi\lesssim x\lesssim d-\xi $. By the same argument, in the right-side uniform region $ d+\xi\lesssim x\lesssim L-\xi $, assuming $ \psi_1=0 $ and $ \psi_2=\sqrt{\rho_2}\mathrm{e}^{\mathrm{i}\theta_2} $, we obtain
	\begin{align}
		u_1=v_1=0,\quad u_2\mathrm{e}^{-\mathrm{i}\theta_2}=-v_2\mathrm{e}^{\mathrm{i}\theta_2}=\cosh k(x-L).
	\end{align}
	We thus obtain
	\begin{align}
		\begin{pmatrix} u_1 \\ u_2 \\ v_1 \\ v_2 \end{pmatrix}=\begin{pmatrix} a \theta(d-x)\mathrm{e}^{\mathrm{i}\theta_1}\cosh k(x+L) \\ b \theta(x-d)\mathrm{e}^{\mathrm{i}\theta_2}\cosh k(x-L) \\ -a \theta(d-x)\mathrm{e}^{-\mathrm{i}\theta_1}\cosh k(x+L) \\ -b \theta(x-d)\mathrm{e}^{-\mathrm{i}\theta_2}\cosh k(x-L) \end{pmatrix}, \label{eq:ripmod0}
	\end{align}
	where the coefficients $ a,b $ are fixed below. This conclusion is more quickly obtained if we assume that $ \epsilon $ is small hence ignorable. \\ 
	\indent Next, by following the procedure (C), we modify the solution (\ref{eq:ripmod0}) to include the zero-mode solution $ w_{\text{rip}} $ [Eq.~(\ref{eq:wrippzd})]. Henceforth we write such modified solution as $ w_{\text{rip}}(k) $. The modified solution must satisfy $ w_{\text{rip}}(0)=w_{\text{rip}} $. From the expression (\ref{eq:ripmod0}), we can conceive the replacement $ \theta(d-x)\mathrm{e}^{\mathrm{i}\theta_1}\rightarrow \psi_1/\sqrt{\rho_1},\ \theta(x-d)\mathrm{e}^{\mathrm{i}\theta_2}\rightarrow \psi_2/\sqrt{\rho_2} $ to include $ w_{\text{rip}} $. Then, we obtain 
	\begin{align}
		w_{\text{rip}}(k) \sim \begin{pmatrix} a'\psi_1\cosh k(x+L) \\ b'\psi_2\cosh k(x-L) \\ -a'\psi_1^*\cosh k(x+L) \\ -b'\psi_2^*\cosh k(x-L) \end{pmatrix}.
	\end{align}
	Here $ a'=a/\sqrt{\rho_1} $ and $ b'=b/\sqrt{\rho_2} $. The ratio of the coefficients $ a',b' $ is fixed by imposing the condition that $ w_{\text{rip}}(k) $ has the same behavior with $ w_{\text{rip}} $ near the DW, that is,
	\begin{align}
		w_{\text{rip}}(k) \simeq w_{\text{rip}} \quad \text{for } x\simeq d.
	\end{align}
	Then, we have
	\begin{align}
		w_{\text{rip}}(k)=\begin{pmatrix} (1-\frac{d}{L})m_1\frac{\cosh k(x+L)}{\cosh k(d+L)}\psi_1 \\[.5ex] -(1+\frac{d}{L})m_2\frac{\cosh k(x-L)}{\cosh k(d-L)}\psi_2 \\[.5ex] -(1-\frac{d}{L})m_1\frac{\cosh k(x+L)}{\cosh k(d+L)}\psi_1^* \\[.5ex] (1+\frac{d}{L})m_2\frac{\cosh k(x-L)}{\cosh k(d-L)}\psi_2^* \end{pmatrix}.
	\end{align}
	It is worth noting that this solution can be used for both Dirichlet and Neumann BCs. Now we solve the Bogoliubov equation by the modified ansatz
	\begin{align}
		\zeta=\alpha w_{\text{rip}}(k)+\beta z_d.
	\end{align}
	If we set $ k=0 $, i.e.,  $ w_{\text{rip}}(k)=w_{\text{rip}}(0)=w_{\text{rip}} $, the ansatz reduces to that in the previous subsection [Eq.~(\ref{eq:twostateansatz})]. By taking the $ \sigma $-inner product between the Bogoliubov equation $ (H_0+M_0k^2)\zeta=\epsilon\zeta $ and $ w_{\text{rip}}(0) $ and $ z_d $, we obtain
	\begin{align}
		\begin{pmatrix}-\epsilon & \frac{Lc_{\text{rip}}^2}{L^2-d^2}\frac{(z_d,w_{\text{rip}}(0))_\sigma}{(z_d,w_{\text{rip}}(k))_\sigma}+k^2\frac{(z_d,M_0z_d)_\sigma}{(z_d,w_{\text{rip}}(k))_\sigma} \\ k^2\frac{(w_{\text{rip}}(0),M_0w_{\text{rip}}(k))_\sigma}{(w_{\text{rip}}(0),z_d)_\sigma}&-\epsilon\end{pmatrix}\begin{pmatrix}\alpha \\ \beta \end{pmatrix}=0
	\end{align}
	where we have used the easily-checked relations $ (w_{\text{rip}}(0),w_{\text{rip}}(k))_\sigma=(z_d,M_0w_{\text{rip}}(k))_\sigma=0 $. The dispersion becomes
	\begin{align}
		\epsilon^2&=A(k)k^2(k^2-k_c^2), \label{eq:ripdispmod}\\
		A(k)&:=\frac{2T_0(w_{\text{rip}}(0),M_0w_{\text{rip}}(k))_\sigma}{(z_d,w_{\text{rip}}(0))_\sigma(z_d,w_{\text{rip}}(k))_\sigma},
	\end{align}
	where $ k_c $ is defined in Eq.~(\ref{eq:ripkc}). 
	Let us evaluate the leading order of $ k $-dependent  $ \sigma $-inner products appearing in $ A(k) $. In fact, the following rough expression is sufficient for this purpose:
	\begin{align}
		|\psi_1|^2= \rho_1\theta(x+L)\theta(d-x), \label{eq:ripcond1rg}\\
		|\psi_2|^2= \rho_2\theta(x-d)\theta(L-x). \label{eq:ripcond2rg}
	\end{align}
	We emphasize that these expressions should not be used to evaluate other $ \sigma $-inner products such as $ 2T_0=(z_d,M_0z_d)_\sigma $. 
	By using Eqs.~(\ref{eq:ripcond1rg}) and (\ref{eq:ripcond2rg}), 
we obtain after some calculations:
	\begin{align}
		&(z_d,w_{\text{rip}}(k))_\sigma=(z_d,w_{\text{rip}}(0))_\sigma \nonumber \\
		&=m_1\rho_1(1-\tfrac{d}{L})+m_2\rho_2(1+\tfrac{d}{L}), \\
		&(w_{\text{rip}}(0),M_0w_{\text{rip}}(k))_\sigma\nonumber \\
		&=m_1\rho_1(1-\tfrac{d}{L})^2\frac{\tanh k(L+d)}{k}+m_2\rho_2(1+\tfrac{d}{L})^2\frac{\tanh k(L-d)}{k}.
	\end{align}
	We thus  obtain
	\begin{align}
		A(k)=& \frac{2T_0}{k[m_1\rho_1(L-d)+m_2\rho_2(L+d)]^2}\times \nonumber \\
		&\quad\left[m_1\rho_1(L-d)^2\tanh k(L+d)\right. \nonumber \\
		&\quad\qquad \left.+m_2\rho_2(L+d)^2\tanh k(L-d)\right]. \label{eq:ripak}
	\end{align}
	This $ A(k) $ has the following two important limiting cases:
	\begin{align}
		A(k) = \begin{cases} A_0 & (kL \ll 1) \\[1ex] \displaystyle \frac{2T_0}{k(m_1\rho_1+m_2\rho_2)} & (L\rightarrow\infty). \end{cases}
	\end{align}
	Here $ A_0 $ is introduced in Eq.~(\ref{eq:ripa0}) and its size-dependence is $ A_0=O(L) $. Correspondingly, the dispersion relation (\ref{eq:ripdispmod}) reduces to
	\begin{align}
		\epsilon^2=\begin{cases} A_0k^2(k^2-k_c^2) & (kL\ll 1) \\[1ex] \displaystyle \frac{2T_0k^3}{m_1\rho_1+m_2\rho_2} & (L\rightarrow\infty). \end{cases}
	\end{align}
	We thus have found that the dispersion relation (\ref{eq:ripdispmod}) includes both  $ \epsilon \sim\sqrt{L}k^2 $ and $ \epsilon \sim k^{3/2} $. Furthermore, the formula (\ref{eq:ripdispmod}) with Eq.~(\ref{eq:ripak}) is valid even for the intermediate region interpolating these two limiting cases.\\
	\indent If the DW is located at the center ($ d=0 $), the expression for the dispersion relation becomes a little simpler:
	\begin{align}
		\epsilon^2=\frac{2T_0}{m_1\rho_1+m_2\rho_2}\frac{\tanh kL}{k}k^2(k^2-k_c^2). \label{eq:rippspmodd0}
	\end{align}
	It includes all three behaviors shown in Table~\ref{ta:rippaprox}:
	\begin{align}
		\epsilon \simeq \sqrt{\frac{2T_0}{m_1\rho_1+m_2\rho_2}}\times \begin{cases} \mathrm{i}\sqrt{L}k_ck, & 0\le k \lesssim O(\mathrm{e}^{-\alpha L/\xi}), \\ \sqrt{L}k^2, & O(\mathrm{e}^{-\alpha L/\xi}) \lesssim k \lesssim O(L^{-1}), \\ k^{3/2}, & O(L^{-1})\lesssim k \lesssim O(\xi^{-1}). \end{cases}
	\end{align}
	The eigenvector is given by
	\begin{align}
		\zeta&=\epsilon w_{\text{rip}}(k)+\frac{k^2A(k)[m_1\rho_1(1-\tfrac{d}{L})+m_2\rho_2(1+\tfrac{d}{L})]}{2T_0}z_d. \label{eq:rippeigengen}
	\end{align}
	If we set $ d=0 $ and take the limit $ L\rightarrow \infty $, we obtain
	\begin{align}
		\zeta \propto \frac{\partial }{\partial x}\begin{pmatrix}\psi_1 \\ \psi_2 \\ \psi_1^* \\ \psi_2^*  \end{pmatrix} - \sqrt{\frac{2T_0}{m_1\rho_1+m_2\rho_2}}k^{1/2} \begin{pmatrix} m_1\psi_1\mathrm{e}^{kx} \\ -m_2\psi_2\mathrm{e}^{-kx} \\ -m_1\psi_1^*\mathrm{e}^{kx} \\ m_2\psi_2^*\mathrm{e}^{-kx} \end{pmatrix},
	\end{align}
	where Eq.~(\ref{eq:zdtowtrans}) is used. It describes the quasiparticle wavefunction of ripplons in the infinite system.

\section{Summary}\label{sec:summary}
In this paper, we have presented 
the analytical formulas interpolating 
the integer dispersion in finite-size systems and 
non-integer dispersion in infinite-size systems 
for the Kelvin modes along a quantized vortex 
and the ripplons on a domain wall in superfluids, 
together with quasiparticle wavefunctions,
and have found a complete agreement between 
our formulas and numerical simulations. 
The derivations of these formulas are supported in a fully analytical way using the techniques constructed in Ref.~\onlinecite{DTMN}.

Finally we give a remark on the criteria for emergence of non-integer dispersion relations. In ferromagnets, 
NGMs  such as
a ripplon on a domain wall \cite{PhysRevLett.113.120403} 
and Kelvon on a skyrmion line \cite{PhysRevLett.112.191804,PhysRevD.90.025010}
have quadratic dispersion relations even for large system sizes.
This is because the zero modes in these systems are normalizable.
On the other hand, in the cases studied in this paper, the zero modes are non-normalizable\cite{DTMN}.

\begin{acknowledgments}
We thank Hiromitsu Takeuchi and Kenichi Kasamatsu for useful comments.
The work of MN is supported in part by Grant-in-Aid for Scientific Research 
No.~25400268
and by the ``Topological Quantum Phenomena'' 
Grant-in-Aid for Scientific Research 
on Innovative Areas (No.~25103720)  
from the Ministry of Education, Culture, Sports, Science and Technology 
(MEXT) of Japan.
The work of MK is supported in part by Grant-in-Aid for Scientific Research No. 26870295,
by Grant-in-Aid for Scientific Research on Innovative Areas ``Fluctuation \& Structure'' (No. 26103519) 
from the Ministry of Education, Culture, Sports, Science and Technology of Japan,
by the JSPS Core-to-Core program ``Non-equilibrium dynamics of soft-matter and information'',
and by the Supercomputer Center, the Institute for Solid State Physics, the University of Tokyo
for the use of the facilities.
\end{acknowledgments}

\appendix
\makeatletter
\renewcommand{\theequation}{%
\thesection\arabic{equation}}
\@addtoreset{equation}{section}
\makeatother
\section{Healing lengths of two-component BECs}\label{app:miscripplon}
	In this appendix we discuss a few fundamental facts on the two-component BEC model such as conservation laws and healing lengths of the DWs. Let us consider an infinite one-dimensional system. The time-dependent GP equation is given by
	\begin{align}
		\mathrm{i}\partial_t\psi_1&=\left(-\mu_1-\tfrac{1}{2m_1}\partial_x^2+2g_{11}|\psi_1|^2+2g_{12}|\psi_2|^2\right)\psi_1, \\
		\mathrm{i}\partial_t\psi_2&=\left(-\mu_2-\tfrac{1}{2m_2}\partial_x^2+2g_{21}|\psi_1|^2+2g_{22}|\psi_2|^2\right)\psi_2.
	\end{align}
	Here we write down the conservation laws. The number conservation laws are
	\begin{align}
		\frac{\partial }{\partial t}|\psi_i|^2+\frac{\partial }{\partial x}\left( \frac{\mathrm{i}(\psi_{ix}^*\psi_i-\psi_i^*\psi_{ix})}{2m_i} \right)=0,\quad i=1,2.
	\end{align}
	The momentum conservation law is given by
	\begin{align}
	&\frac{\partial }{\partial t}\left[ \sum_{i=1,2}\frac{\mathrm{i}(\psi_{ix}^*\psi_i-\psi_i^*\psi_{ix})}{2} \right]\nonumber \\
	&+\frac{\partial }{\partial x}\left[ \sum_{i=1,2}\left(\frac{\mathrm{i}(\psi_i^*\psi_{it}-\psi_{it}^*\psi_i)}{2}+\frac{|\psi_{ix}|^2}{2m_i}+\mu_i|\psi_i|^2\right)\right. \nonumber \\
	&\qquad\qquad\left.-\sum_{i,j=1,2}g_{ij}|\psi_i|^2|\psi_j|^2 \right]=0.
	\end{align}
	We omit the energy conservation law because it does not give a new integration constant for a time-independent solution. From these conservation laws, for the stationary solution $ \psi_{1t}=\psi_{2t}=0 $, we have the following integration constants:
	\begin{align}
		&j_i=\frac{\mathrm{i}(\psi_{ix}^*\psi_i-\psi_i^*\psi_{ix})}{2m_i},\quad i=1,2, \\
		&j_{\text{mom}}=\sum_{i=1,2}\left(\frac{|\psi_{ix}|^2}{2m_i}+\mu_i|\psi_i|^2\right)-\sum_{i,j=1,2}g_{ij}|\psi_i|^2|\psi_j|^2. \label{eq:jmom001}
	\end{align}
	If $ \psi_1,\psi_2 $ are real, $ j_1=j_2=0 $, and hence  $ j_{\text{mom}} $ is the only non-trivial constant.\\
	\indent Let us consider the DW solution having the following asymptotic form:
	\begin{align}
		\psi_1\rightarrow\begin{cases}0 & (x\rightarrow+\infty) \\ \sqrt{\rho_1} & (x\rightarrow-\infty), \end{cases}\quad \psi_2\rightarrow\begin{cases}\sqrt{\rho_2} & (x\rightarrow+\infty) \\ 0 & (x\rightarrow-\infty). \end{cases}
	\end{align}
	In order for this asymptotic form to become the solution of the GP equation, the values of the chemical potentials should be fixed as
	\begin{align}
		\mu_i=2g_{ii}\rho_i,\quad i=1,2.
	\end{align}
	Furthermore, from the  $ x $-independence of the momentum current density (\ref{eq:jmom001}), we obtain the relation
	\begin{align}
		j_{\text{mom}}=g_{11}\rho_1^2=g_{22}\rho_2^2, \label{eq:ripplonap2}
	\end{align}
	which is the same with Eq.~(\ref{eq:ripplonp2}). Thus, $ \rho_1 $ and $ \rho_2 $ cannot be chosen independently. We also note that the meaning of the parameter $ p $ is, in fact, the square root of the momentum current: $ j_{\text{mom}}=p^2 $. \\
	\indent Let us introduce four kinds of healing lengths. We first consider the situation such that only $ \psi_1 $ exists. In this case Eq.~(\ref{eq:jmom001}) reduces to
	\begin{align}
		\psi_{1x}^2=2m_1g_{11}(\psi_1^2-\rho_1)^2,
	\end{align}
	and a solution is given by the well-known dark soliton solution:
	\begin{align}
		\psi_1&=\sqrt{\rho_1}\tanh\frac{\kappa_1 x}{2},\\
		\kappa_1&:=2\sqrt{2g_{11}m_1\rho_1}. \label{eq:apphl1}
	\end{align}
	This $ \kappa_1 $ describes the inverse of the healing length for the one-component system. In the same way, we obtain that for $ \psi_2 $: 
	\begin{align}
		\kappa_2:=2\sqrt{2g_{22}m_2\rho_2}.
	\end{align}
	Next let us consider the decay rate of $ \psi_1 $ on the right side of the DW, where $ \psi_2 $ is dominant. Assuming  $ \psi_1 $ is small and $ \psi_2 \simeq \sqrt{\rho_2} $, the GP equation can be approximated as
	\begin{align}
		\tfrac{-\partial_x^2\psi_1}{2m_1}+2(g_{12}\rho_2-g_{11}\rho_1)\psi_1=0,
	\end{align}
	where the nonlinear term is ignored with assuming small $ \psi_1 $. Then,
	\begin{align}
		&\psi_1 \propto \mathrm{e}^{-\kappa_{\text{DW1}}x},\\
		&\kappa_{\text{DW1}}:=2\sqrt{m_1\rho_2(g_{12}-\sqrt{g_{11}g_{22}})}.
	\end{align}
	This $ \kappa_{\text{DW1}} $ represents the decay rate. Here, we have used Eq.~(\ref{eq:ripplonap2}) to obtain $ g_{12}\rho_2-g_{11}\rho_1=\rho_2(g_{12}-\sqrt{g_{11}g_{22}}) $. By the same calculation, on the left side of the DW, we can show
	\begin{align}
		&\psi_2 \propto \mathrm{e}^{\kappa_{\text{DW2}}x},\\
		&\kappa_{\text{DW2}}:=2\sqrt{m_2\rho_1(g_{12}-\sqrt{g_{11}g_{22}})}. \label{eq:apphldw2}
	\end{align}
	Summarizing, we have obtained four inverse healing lengths, $ \kappa_1,\ \kappa_2, \ \kappa_{\text{DW1}}, $ and $ \kappa_{\text{DW2}} $. Thus, the term ``typical healing length $ \xi $'' used in Secs.~\ref{sec:mainresult} and \ref{sec:ripplon} precisely means the largest one among these four lengths, i.e.,
	\begin{align}
		\xi = \max(\kappa_1^{-1},\kappa_2^{-1},\kappa_{\text{DW1}}^{-1},\kappa_{\text{DW2}}^{-1}).
	\end{align}
\section{Evaluation of $ k_c $ for the case of $ g_{12}=\infty $}\label{app:evaluatemuderiv}
	In this appendix, we focus on the system with $ g_{12}=+\infty $, in which two condensates $ \psi_1,\psi_2 $ are completely decoupled and hence the GP equation reduces that of a single-component BEC. We want to find the leading  $ L $-dependence of $ k_c $ [Eq.~(\ref{eq:ripkc})], the maximum wavenumber such that the dispersion relation of ripplons becomes complex-valued, in other words, the maximum wavenumber of unstable modes. For simplicity,  we only concentrate on the case where the physical parameters of two BECs are symmetric, i.~e.,  $ g_{11}=g_{22}=1,\ 2m_1=2m_2=1,\ N_1=N_2 $. In this case, $ \mu_{1d}=-\mu_{2d} $ holds by symmetry.\\
	\indent Both $ \psi_1 $ and $ \psi_2 $ satisfy the single-component GP equation
	\begin{align}
		-\psi''-\mu\psi+2\psi^3=0,
	\end{align}
	and the general solution is given by
	\begin{align}
		\psi(x;\bar{\rho},m)&=\sqrt{\tfrac{\bar{\rho}m}{Q(m)}}\operatorname{sn}\left(\!\left.\sqrt{\tfrac{\bar{\rho}}{Q(m)}}x\right|m \right),\label{eq:appg12psi} \\
		Q(m)&=1-\frac{E(m)}{K(m)},\\
		\mu&=\frac{(1+m)\bar{\rho}}{Q(m)},
	\end{align}
	where  $ K $ and $ E $ are the complete elliptic integral of the first and second kind, respectively. Here and hereafter, we use Mathematica's notations for the elliptic integrals/functions unless otherwise noted. The solution (\ref{eq:appg12psi}) is characterized by two parameters $ m $ and $ \bar{\rho} $. The former is an elliptic parameter and satisfy $ 0< m\le 1 $. The latter has the physical meaning of the averaged particle number density:
	\begin{align}
		\frac{1}{K(m)\sqrt{Q(m)/\bar{\rho}}}\int_0^{K(m)\sqrt{Q(m)/\bar{\rho}}}\mathrm{d}x|\psi|^2=\bar{\rho}.
	\end{align}
	The energy per particle can be calculated as
	\begin{align}
		\frac{E}{N}=\frac{\int_0^{K(m)\sqrt{Q(m)/\bar{\rho}}}\mathrm{d}x(|\psi'|^2+|\psi|^4)}{\int_0^{K(m)\sqrt{Q(m)/\bar{\rho}}}\mathrm{d}x|\psi|^2}=\frac{[m+(1+m)Q(m)]\bar{\rho}}{3Q(m)^2}.
	\end{align}
	Henceforth, we write the physical parameters of $ \psi_i$'s ($i=1,2$) as  $ \ m_i,\ \bar{\rho}_i,\ \mu_i,\ N_i,\ E_i, $ and so on. \\
	\indent If we use the Dirichlet  BC ($ \psi_i=0 $ at the boundary), the profiles of $ \psi_i $'s are given by the sn function with one-half of a period. If we use the Neumann BC ($\psi_i'=0$ at the boundary), the profiles of $ \psi_i $'s are given by the sn function with one-quarter of a period. Therefore, the length $ L_i $ of the region that $ \psi_i $ occupies is given by
	\begin{align}
		L_i&=\alpha K(m_i)\sqrt{\tfrac{Q(m_i)}{\bar{\rho}_i}}, \\
		\alpha&=\begin{cases} 1 & \text{(the Neumann BC)} \\ 2 & \text{(the Dirichlet BC)}. \end{cases}
	\end{align}
	Needless to say,  $ L_1 $ and $ L_2 $ are not independent and satisfy $ L_1+L_2=2L $. \\
	\indent Since we want to solve the energy minimization problem with respect to $ L_1 $ under the condition that $ N_1,N_2,L $ are fixed, we change the independent variables from $ m_1,\bar{\rho}_1,m_2,\bar{\rho}_2 $  to $ N_1,L_1,N_2,L_2 $. Their relations are given by
	\begin{align}
		\bar{\rho}_i&=\frac{N_i}{L_i}, \\
		\frac{L_iN_i}{\alpha^2}&=K(m_i)^2Q(m_i).
	\end{align}
	Thus, in order to move on to the description by $ N_i $ and $ L_i $, we need an inverse function of $ K(m)^2Q(m) $. Though the exact inverse function cannot be written down in a closed form, if $ m\simeq 1 $ (i.e., if sn is almost tanh), we obtain the following asymptotic expansion:
	\begin{align}
		x&=K(m)^2Q(m) \\
		\leftrightarrow\quad m&=1-16\mathrm{e}^{-y}+128\mathrm{e}^{-2y}+\dotsb,\quad y:=1+\sqrt{1+4x}. \label{eq:ripappk2qinv}
	\end{align}
	The expansion (\ref{eq:ripappk2qinv}) can be obtained by using the formulas
	\begin{align}
		K(1-16\delta)&=-\frac{1}{2}\log\delta-2\delta(2+\log\delta)+O(\delta^2\log\delta),\\
		E(1-16\delta)&=1-4\delta(1+\log\delta)+O(\delta^2\log\delta)
	\end{align}
	and solving the equation $ x=K^2Q=K^2-KE $ w.r.t $ \delta $ iteratively. When Eq.~(\ref{eq:ripappk2qinv}) is applicable,  $ K(m) $ and $ Q(m) $ are given by
	\begin{align}
		K(m)&=y\left( \tfrac{1}{2}+2\mathrm{e}^{-y}-16\mathrm{e}^{-2y}+\dotsb \right),\\
		Q(m)&=\frac{x}{K(m)^2}=\left( 1-\tfrac{2}{y} \right)\left( 1-8\mathrm{e}^{-y}+112\mathrm{e}^{-2y}+\dotsb \right).
	\end{align}
	By using them, the chemical potential and the energy for $ \psi_i $ are written as a function of $ (L_i,N_i) $: 
	\begin{align}
		\mu_i&=\frac{2N_i}{L_i}\left( 1+\tfrac{2}{y} \right)\left( 1-48\mathrm{e}^{-2y}+\dotsb \right),\\
		E_i&=\frac{N_i^2}{L_i}\left[ \left( 1+\tfrac{8}{3y} \right)-\tfrac{64}{3}\left( 4+\tfrac{13}{y} \right)\mathrm{e}^{-2y} \right],\\
		y:&=1+\sqrt{1+\tfrac{L_iN_i}{\alpha^2}}
	\end{align}
	Here, the terms of order $O(y^{-a}\mathrm{e}^{-2by})$ with $ a\ge2 $ or $ b\ge 2 $ are ignored.\\
	\indent Now, let us write
	\begin{align}
		L_1=L+\delta L,\ L_2=L-\delta L,\\
		N_1=L\rho_0+\delta N,\ N_2=L\rho_0-\delta N.
	\end{align}
	where $ \rho_0=\frac{N_1+N_2}{2L} $ is the average of the total particle number density. Let us minimize
	\begin{align}
		E_{\text{total}}=E_1+E_2
	\end{align}
	with respect to $ \delta L $ under the constraint that $ L,\rho_0, $ and $ \delta N $ are fixed. If $ \delta N=0 \ \leftrightarrow N_1=N_2 $, we immediately obtain a trivial solution $ \delta L=0 $. Let us find  $ \delta L $ for the non-zero imbalance $ \delta N\ne0 $. After a little tedious calculation, we obtain
	\begin{align}
		&\frac{\partial E_{\text{total}}}{\partial \delta L}=0 \ \leftrightarrow \nonumber \\
		&\delta L\simeq \delta N\left[ \tfrac{1}{\rho_0}\Bigl( 1-\tfrac{\alpha}{L\sqrt{\rho_0}} \Bigr)+\tfrac{1024L^2\rho_0\mathrm{e}^{-2-\frac{4L\sqrt{\rho_0}}{\alpha}}}{3\alpha^2}\Bigl( 1-\tfrac{9\alpha}{8L\sqrt{\rho_0}} \Bigr) \right]\nonumber \\
		&\qquad\qquad +O(\mathrm{e}^{-\frac{8L\sqrt{\rho_0}}{\alpha}},\delta N^2),
	\end{align}
	where  $ O(L^{-2}) $ terms are ignored in each parenthesis. By using this $ \delta L $, up to the same approximation, $ \mu_1 $ can be written as
	\begin{align}
		\mu_1=2\rho_0\Bigl( 1+\tfrac{\alpha}{L\sqrt{\rho_0}} \Bigr)-\delta N\tfrac{2048\rho_0L\mathrm{e}^{-2-\frac{4L\sqrt{\rho_0}}{\alpha}}}{3\alpha^2}\Bigl( 1-\tfrac{3\alpha}{16L\sqrt{\rho_0}} \Bigr).
	\end{align}
	In the present calculation, recalling that we have set $ g_{11}=g_{22}=1 $, the parameters $ p $ and $ d $ introduced in Subsec.~\ref{subsec:ripGS} are
	\begin{align}
		p=\rho_0,\ d=\frac{\delta N}{2\rho_0}.
	\end{align}
	Thus, the $ d $-derivative of $ \mu_1 $ up to leading order is given by
	\begin{align}
		\mu_{1d}=\frac{\partial \mu_1}{\partial d} \simeq -\frac{4096\rho_0^2L}{3\alpha^2}\mathrm{e}^{-2-\frac{4L\sqrt{\rho_0}}{\alpha}}.
	\end{align} 
	It is obviously negative: $ \mu_{1d}<0 $. By ignoring the $ O(1) $ numerical factor, the main  $ L $-dependence can be given by
	\begin{align}
		\mu_{1d} \sim \begin{cases} -L\mathrm{e}^{-4L\sqrt{\rho_0}} & (\alpha=1; \text{ Neumann BC}) \\ -L\mathrm{e}^{-2L\sqrt{\rho_0}} & (\alpha=2; \text{ Dirichlet BC}).  \end{cases}
	\end{align} 
	Since $ k_c \propto \sqrt{-\mu_{1d}} $ [Eq.~(\ref{eq:ripkcsym})], we also obtain 
	\begin{align}
		k_c \sim \begin{cases} \sqrt{L}\mathrm{e}^{-2L\sqrt{\rho_0}} & (\alpha=1; \text{ Neumann BC}) \\ \sqrt{L}\mathrm{e}^{-L\sqrt{\rho_0}} & (\alpha=2; \text{ Dirichlet BC}).  \end{cases}
	\end{align} 
	We thus have proved the behavior in Eq.~(\ref{eq:muidbehav}). \\
	\indent Though this result is rigorously applicable only for the special case $ g_{12}=\infty $, the numerical results suggest that the above behavior is also true for finite $ g_{12} $ if we modify the exponential factor as $ \mathrm{e}^{-\frac{2L\sqrt{\rho_0}}{\alpha}} \rightarrow \mathrm{e}^{-\nu\frac{2L\sqrt{\rho_0}}{\alpha}} $, where $ \nu \sim 1 $ is a numerical fitting parameter. See Fig. \ref{fig:ripkc}. Thus, we can say that $ k_c $ is always exponentially small. \\
	\indent The above result suggests that the Neumann BC can suppress unstable modes more strongly than the Dirichlet BC. For example, if we set $ L=12 $ and $ \rho_0=1 $, then $ k_c \sim 10^{-5} $ for the Dirichlet BC and $ k_c\sim10^{-10} $ for the Neumann BC. This means that the typical eigenenergies of the complex-valued regions are given by $ |\epsilon|\sim O(k_c^2)\sim 10^{-10} $ for the Dirichlet BC and  $ |\epsilon|\sim O(k_c^2)\sim 10^{-20} $ for the Neumann BC. While the former might be numerically seen, the latter is impossible to detect in the usual precision. Therefore, the Neumann BC is a powerful tool if one is interested in the infinite-size physics and wants to ignore finite-size effects, though sometimes this BC is not physically realistic. This observation is consistent with the previous numerical study performed in the Neumann BC in Ref.~\onlinecite{PhysRevA.88.043612}, where no unstable mode was found numerically for large $ L $.

%

\end{document}